\documentclass[prb,twocolumn,showpacs,superscriptaddress]{revtex4}
\usepackage[latin1]{inputenc}
\usepackage{graphicx}
\usepackage{amssymb}
\usepackage{amsmath}
\usepackage{xspace}
\usepackage{dcolumn}
\usepackage{bm}

\newfont{\tensy}{cmsy10}
\newcommand{\chem}[1]{{$\fontdimen16\tensy=3.0pt
    \fontdimen17\tensy=3.0pt \mathrm{#1}$}}

\renewcommand{\Im}[0]{\text{Im}\,}

\newcommand{\etal}[0]{{\it et al.}\@\xspace}
\newcommand{\ie}[0]{i.e.\@\xspace}
\newcommand{\eg}[0]{e.g.\@\xspace}

\newcommand{\rmi}{\mathrm{i}}

\newcommand{\UP}[0]{\uparrow}
\newcommand{\DO}[0]{\downarrow}

\newcommand{\sign}{\las\text{sign}\ras}
\newcommand{\ek}[0]{\varepsilon_k}
\newcommand{\om}[0]{\omega}
\newcommand{\Ek}{\overline{E}_\text{k}}
\newcommand{\Ep}{E_\text{P}}
\newcommand{\omb}[0]{\overline{\omega}_0}

\newcommand{\Z}[0]{\mathcal{Z}}
\newcommand{\D}[0]{\mathcal{D}}
\newcommand{\tr}[0]{\text{Tr}\,}
\newcommand{\dtau}{\Delta\tau}

\newcommand{\nag}{\phantom{\dag}}
\newcommand{\op}{\hat{p}}
\newcommand{\ox}{\hat{x}}
\newcommand{\on}{\hat{n}}
\newcommand{\wb}{w_\text{b}}
\newcommand{\wf}{w_\text{f}}

\newcommand{\rD}{\text{D}}
\newcommand{\oB}[0]{\hat{B}}
\newcommand{\V}{\mathcal{V}}
\renewcommand{\H}{\mathcal{H}}

\newcommand{\las}[0]{\langle}
\newcommand{\ras}[0]{\rangle}
\newcommand{\la}[0]{\left\las}
\newcommand{\ra}[0]{\right\ras}
\newcommand{\ket}[1]{\left|#1\ra}
\newcommand{\bra}[1]{\la#1\right|}
\newcommand{\sket}[1]{|#1\ras}  
\newcommand{\sbra}[1]{\las#1|} 

\begin{document}


\title{Photoemission spectra of many-polaron systems}

\author{M.~Hohenadler}\email{hohenadler@itp.tugraz.at}
\author{D.~Neuber}
\author{W.~von der Linden}
\affiliation{%
  Institute for Theoretical and Computational Physics, TU Graz,
  8010 Graz, Austria}
\author{G.~Wellein}
\affiliation{%
  Computing Center, University Erlangen, 91058 Erlangen,
  Germany}
\author{J.~Loos}
\affiliation{%
  Institute of Physics, Czech Academy of Sciences, 16200 Prague, Czech Republic}
\author{H.~Fehske}\email{fehske@physik.uni-greifswald.de}
\affiliation{%
  Institute for Physics, Ernst-Moritz-Arndt University Greifswald, 17487
  Greifswald, Germany}

\begin{abstract}
  The cross over from low to high carrier densities in a many-polaron system
  is studied in the framework of the one-dimensional spinless Holstein model,
  using unbiased numerical methods. Combining a novel quantum Monte Carlo
  approach and exact diagonalization, accurate results for the single-particle
  spectrum and the electronic kinetic energy on fairly large systems are
  obtained.  A detailed investigation of the quality of the Monte Carlo data
  is presented. In the physically most important adiabatic intermediate
  electron-phonon coupling regime, for which no analytical results are
  available, we observe a dissociation of polarons with increasing band
  filling, leading to normal metallic behavior, while for parameters favoring
  small polarons, no such density-driven changes occur. The present work
  points towards the inadequacy of single-polaron theories for a number of
  polaronic materials such as the manganites.
\end{abstract}


\maketitle

%
%
%
\section{\label{sec:introduction}Introduction}
%
%
%

In recent years, it has become widely accepted that electron-phonon (EP)
interaction plays a crucial role in systems with strong electronic
correlations, such as high-temperature superconducting
cuprates\cite{BYMdLBi92} or colossal magnetoresistive
manganites.\cite{David_AiP} A large amount of the available data on optical
and transport properties has been interpreted in terms of polaronic states.
In view of the complex underlying physics, however, the application of
single-polaron theories seems to be problematic.  On the other hand, a
reliable theory of strongly coupled EP systems with finite carrier densities
is not available yet. While limiting cases, such as weak coupling or large
phonon frequencies---compared to the electronic bandwidth---can to some degree be
described theoretically, especially the adiabatic regime of small phonon
frequencies at intermediate-to-strong EP
coupling constitutes a long-standing, open problem.  Since analytical results
can only be obtained at the expense of uncontrolled approximations, progress
toward an understanding of many-polaron systems requires unbiased numerical
methods.  Owing to the availability of high performance computers, the latter
can yield insight into the properties of basic microscopic models
such as the spinless Holstein model considered here. Nevertheless, apart from
work on the half-filled band case, where the physics is dominated by a
cross over from a Luttinger liquid to a Peierls state, the many-polaron
problem has received little attention in the past.

Motivated by this situation, in the present work, we address the important
issue of the effect of carrier density on the character of the quasiparticles
of the system. While for very strong couplings no significant changes are
expected due to the existence of rather independent small (self-trapped)
polarons with negligible residual interaction, a density-driven cross over
from a state with large polarons to a metal with weakly dressed electrons may
occur in the intermediate coupling regime. This problem has recently been
investigated experimentally in terms of optical measurements on
\chem{La_{2/3}(Sr/Ca)_{1/3}MnO_3} films.\cite{HaMaDeLoKo04,HaMaLoKo04} While
small polarons have been identified as the charge carriers in the
low-temperature metallic phase of the Ca doped system, the weaker
electron-lattice coupling in the Sr doped system gives rise to
large-polaron-like behavior of the optical conductivity. The effect of the
finite carrier density on such extended polaronic quasiparticles has been
analyzed\cite{HaMaDeLoKo04,HaMaLoKo04} using a recently developed
weak-coupling theory.\cite{TeDe01}

For the case of the Holstein model, the abovementioned density-driven
transition from small to large polarons and finally to weakly EP-dressed
electrons is expected to be possible only in one dimension (1D), where large
polarons extending over more than one lattice site exist at weak coupling. 
Existing work indicates that in higher dimensions, the electrons
remain quasifree at weak coupling, and become self-trapped above a critical value
of the EP coupling strength.\cite{dRLa82,dRLa83,KaMa93,FeRoWeMi95} By
contrast, approximate variational calculations suggest that large polarons
may also exist in $D>1$.\cite{RoBrLi99III,CadFIa99,CaFiIa01}
The situation is different for Fr\"{o}hlich-type models\cite{Fr54,AlKo99}
with long-range EP interaction, in which large polaron states exist even for
strong coupling and in higher dimensions. 

Here, we make use of quantum Monte Carlo (QMC) and
exact diagonalization (ED) methods, and calculate the photoemission spectra
for finite clusters at zero and finite temperature. Due to the limitations of
existing algorithms, we develop a new grand-canonical QMC approach which is
free of autocorrelations and therefore enables us to study small phonon
frequencies and low temperatures. 

The organization of this paper is as follows. In Sec.~\ref{sec:model}, we
present the model and discuss previous work, while Sec.~\ref{sec:methods}
contains details about the methods we employed. Section~\ref{sec:results}
gives a discussion of our numerical data, and finally, we conclude in
Sec.~\ref{sec:conclusions}. The new QMC technique is described in the
Appendix. 

%
%
%
\section{\label{sec:model}Model}
%
%
%

The spinless Holstein model is defined by the Hamiltonian
\begin{equation}\label{eq:model:H}
  H
  =
  -t \sum_{\las i,j\ras} c^\dag_i c^{\nag}_j
  +\om_0\sum_i b^\dag_i b^{\nag}_i
  -g\om_0 \sum_i \on_i (b^\dag_i + b^{\nag}_i)
  \,,
\end{equation}
where $c^\dag_i$ ($c^{\nag}_i$) creates (annihilates) a spinless fermion at
lattice site $i$, $b^\dag_i$ ($b^{\nag}_i$) creates (annihilates) a phonon of
energy $\om_0$ ($\hbar=1$) at site $i$, and $\on_i=c^\dag_i c^{\nag}_i$. 
While the first term describes hopping processes between neighboring lattice
sites $\las i,j\ras$, the second term corresponds to the elastic and kinetic
energy of the phonons.  Finally, the last term in Eq.~(\ref{eq:model:H})
represents a local coupling of the lattice displacement to the electron
occupation $n_i=0,\,1$. 

The parameters of the model are the hopping integral $t$, the phonon energy
$\om_0$, and the coupling constant $g$. The latter is related to the polaron
binding energy $\Ep$ via $g = \sqrt{\Ep/\om_0}$. By defining the
dimensionless quantities $\lambda = \Ep/2t\rD$, and $\omb=\om_0/t$, we are
left with two independent parameters. In the sequel, we express all energies
in units of $t$. The lattice constant is taken to be unity, and periodic
boundary conditions are applied. 

The Holstein Hamiltonian~(\ref{eq:model:H}) represents a generic model for a
many-polaron system. Due to the restriction to spinless fermions, on-site
bipolaron formation is suppressed.\cite{BoKaTr00} 

Quantum fluctuations of the lattice enable the electrons to hop even in the
strong-coupling regime, while in models with classical phonons translation
symmetry is broken, so that no itinerant polaron states exist.  Along this
line, the half-filled spinless Holstein model has been studied in one
dimension using QMC,\cite{HiFr82,HiFr83II,MKHaMu96} ED,\cite{WeFe98}
DMRG\cite{BuMKHa98} and variational
methods.\cite{ZhFeAv89,FeCiPa90,ZhAv98,WaZhAv00,PeCaDFIaMRVe03} Similar to the
spinfull model at half filling, the spinless model with one electron per two
lattice sites displays a quantum phase transition from a metallic state to a
Peierls insulating state with charge-density-wave (CDW) order
at $T=0$. While the
insulating CDW state of the spinless model is stable for any $\lambda>0$ in
the adiabatic limit $\omb=0$, it can be destroyed by quantum phonon
fluctuations for $\omb>0$.\cite{BuMKHa98} Capone \etal\cite{CaGrSt99} used ED
to study the cross over from a single polaron to a many-electron system away
from half filling. Although their work is actually for the Holstein-Hubbard
model, results are very similar to the spinless Holstein model in the limit
of large $U$ considered in Ref.~\onlinecite{CaGrSt99}. The authors found that
the critical coupling for polaron formation is unaffected by the electron
density in the antiadiabatic regime. Finally, we would like to point out
previous analytical efforts to take into account squeezing and correlation
effects in many-polaron systems.\cite{ZhFeAv89,FeCiPa90,FeIhLoTrBu94}

%
%
%
\section{\label{sec:methods}Methods}
%
%
%

In this work we use QMC, and ED in combination with the kernel polynomial
method (KPM), to calculate, in particular, single-particle spectra. 
These two approaches are characterized by different advantages and
shortcomings.  ED in the form used here yields exact spectra at $T=0$ with
extremely high energy resolution. However, even for relatively small clusters
of ten sites---the size used here---this method requires parallel
supercomputers to handle the enormous Hilbert space (matrix dimension
$\approx10^7$\,--\,$10^{10}$). Consequently, the momentum resolution,
determined by the cluster size, is rather crude. 

In contrast, the grand-canonical QMC simulations can be performed on personal
computers, and for larger clusters. The drawback is that (a) the method is
limited to finite temperatures, and (b) the calculation of spectral
properties requires the analytic continuation from imaginary time to real
frequencies, which is an ill-conditioned problem. Therefore, regularization
techniques such as the maximum entropy method (MEM) used here have to be
applied, which offer only limited energy resolution and may introduce
uncontrolled errors. For this reason, we will also present a direct
comparison of ED and QMC results. 

\subsection{Quantum Monte Carlo}\label{sec:methods:QMC}

In applying existing QMC methods for the Holstein model to the parameter
regime considered here, one faces extremely long autocorrelations times, as
well as large statistical errors.\cite{HoEvvdL03} This makes it very
difficult to obtain accurate results for spectral properties, due to the use
of the MEM. To overcome these problems, the algorithm presented in the
Appendix is based on the canonical Lang-Firsov
transformation.\cite{LangFirsov} The latter explicitly contains the shift of
the lattice equilibrium position in the presence of an electron, which is
favorable for the QMC simulations.  Moreover, the absence of a direct
coupling between the electron density and the lattice displacements in the
transformed Hamiltonian~(\ref{eq:app:Hspinless}) permits exact, uncorrelated
sampling of the phonon degrees of freedom (momenta) in terms of principal
components. 

Here we have mainly used a value of $\beta t=8$ for the inverse temperature
$\beta=1/k_\text{B} T$. For the critical parameters studied below, accurate
simulations at even lower temperatures become very demanding due to a minus-sign
problem (Sec.~\ref{sec:app:sign_results}). Nevertheless, we shall present
results for the density of states at $\beta t=10$ and half filling but for a
reduced cluster size.  Note that already for $\beta t=8$ we have
$k_\text{B}T/t=0.125<\omb$ for the values $\omb\geq0.4$ used in the sequel,
so that direct thermal excitation of phonons is expected to be rather
unimportant. 

For the Trotter decomposition, we chose $\dtau=0.05$ (see Appendix) for
static observables. Owing to the larger numerical effort associated with the
calculation of dynamical properties, for the latter we
use $\dtau=0.1$---the larger error being small compared to the
uncertainties introduced by the MEM inversion.  The Trotter error in the
Holstein model becomes important mainly for large phonon frequencies and very
strong EP interaction.\cite{Ko97,HoEvvdL03} Since we do not consider this
regime here, the abovementioned values of $\dtau$ yield satisfactorily small
systematic errors. 

Away from half filling, the electron density has to be adjusted to the values
of interest by varying the chemical potential $\mu$. Due to the computational
effort associated with this trial-and-error procedure, the actual band
filling $n'$, say, is usually only very close to a desired value, \eg,
$n=0.4$, but not exactly the same.  Furthermore, within the QMC simulations,
the filling can only be determined up to a statistical error $\Delta n'$,
similar to other observables. For the results presented here, the relative
deviation $(n-n')/n$ of the actual value in the QMC simulations from the
value $n$ reported is always smaller than $5\times10^{-3}$. The same is true
for the relative statistical error $\Delta n'/n$. 

To perform the continuation of the one-electron Green function from imaginary
time to real frequencies, we use the MEM (for a review see, \eg,
Ref.~\onlinecite{JaGu96}). The algorithm employed here incorporates an improved evidence
approximation,\cite{vdLPrDo99} which is less susceptible to overfitting---resulting in 
peaks whose position, weights and even existence are not clear---than
the conventional evidence approximation.\cite{JaGu96} On the other hand, groups
of narrowly spaced peaks are represented by an envelope if the
accuracy of the QMC data is not sufficient. 

The proposal of an alternative evidence approximation---leading in general
to stronger regularization and, consequently, lower energy resolution than
the conventional approximation---may not seem advantageous at first. 
However, considering the great variety of different spectra compatible with
the QMC data within the error bars due to the extremely
ill-conditioned analytical continuation problem, we argue that it
rather corresponds to choosing the ``most uninformative'' solution in the
spirit of the maximum entropy principle. In fact, we found that the
alternative evidence approximation yields much more consistent results
than the ``classic'' evidence approximation\cite{JaGu96} when we subjected
artificial test data to different noise while keeping the signal-to-noise
ratio fixed. A detailed comparison of different regularization schemes available
will be presented elsewhere. 

In order to scrutinize the resulting spectra, the following test has been
performed. Following the idea of Silver \etal,\cite{SiSiGu90} the results
from the MEM applied to the original QMC data $G(k,\tau)$ have been taken as
an ``exact'' spectral function $A(k,\om)$, from which mock data
$\tilde{G}(k,\tau)$ for the imaginary-time Green function has been created by
using Eq.~(\ref{eq:app:maxent}) and adding Gaussian random noise of absolute
size $10^{-3}$\,--\,$10^{-4}$, comparable to the statistical errors of the
QMC results. Again using the MEM to invert these data, the thereby obtained
spectral function $\tilde{A}(k,\om)$ allows to check the stability of the MEM
result $A(k,\om)$ with respect to statistical errors. We find that the
features in our results for $A(k,\om)$ remain stable if $G(k,\om)$ varies
within the statistical errors, so that the data presented below are
satisfactorily reliable apart from the aforementioned broadening and the
limited energy resolution. 

While our method is completely free of autocorrelations between subsequent
phonon configurations, results for the imaginary-time Green function at
different times are usually not statistically independent.\cite{vdLPrHa96}
However, in the present case, owing to the exact sampling and the large
number of measurements, these correlations are very small and have been
neglected since they have negligible influence on the results.  An
additional, stringent test of the QMC results consists of a comparison to ED
data which will be presented in Sec.~\ref{sec:results:intermediate}. 

Finally, we note that all QMC simulations have been run on personal computers
(\eg, Intel XEON 2600 MHz), with CPU times varying between several hours
and a couple of days.

\subsection{Exact diagonalization}\label{sec:methods:ED}

The combination of ED and KPM has been described in detail in
Ref.~\onlinecite{BaWeFe98}. In this work, we are mainly interested in the
single-particle spectral functions
\begin{eqnarray}\label{eq:methods:Apm}
  &&\begin{split}\nonumber
    A^+(k,\om)
    = 
    \sum_l
    |
    &
    \sbra{\Psi_{l,k}^{(N_\text{e}+1)}} c^\dag_{k-q} \sket{\Psi_{0,q}^{(N_\text{e})}}
    |^2\,
    \\
    &
    \times
    \delta[\om - (E_{l,k}^{(N_\text{e}+1)} - E_{0,q}^{(N_\text{e})})]
    \,,
  \end{split}
  \\
  &&\begin{split}
    A^-(k,\om)
    = 
    \sum_l
    |
    &
    \sbra{\Psi_{l,k}^{(N_\text{e}-1)}}
    c^{\nag}_{q-k}
    \sket{\Psi_{0,q}^{(N_\text{e})}}
    |^2\,
    \\
    &
    \times
    \delta[\om + (E_{l,k}^{(N_\text{e}-1)} - E_{0,q}^{(N_\text{e})})]
    \,,
  \end{split}
\end{eqnarray}
where $\sket{\Psi^{(N_\text{e})}_{l,k}}$ denotes the $l$th eigenstate with
$N_\text{e}$ electrons, momentum $k$ and energy $E_{l,k}^{(N_\text{e})}$. The
cluster size has been fixed at $N=10$, with as many as 15 dynamical
phonons and resulting truncation errors $<10^{-6}$ for the ground-state
energy.\cite{BaWeFe98}  Additionally, the convergence with the number of phonons has been
monitored by means of the phonon distribution function.\cite{BaWeFe98} It is
important to point out that such a small number of dynamical phonons only
yields accurate results for the parameters considered here if the symmetric
phonon mode with $q=0$ is separated from the Hamiltonian analytically before
the diagonalization.\cite{Robin97,SyHuBeWeFe04} In the present case, this
phonon mode---corresponding to a homogeneous shift of the oscillator
equilibrium positions---together with that for $q=\pi$ represents the
dominant contribution to phonon excitations, so that the separation is very
efficient at reducing the computational effort. The largest matrix dimension
in the present calculations was about $3\times 10^8$,
running on six compute nodes of a Hitachi SR8000-F1.  The energy
resolution of the KPM spectral function with 512 moments has been enhanced by
a factor of 8 by the MEM,\cite{BaWeFe98} resulting in
a significantly better resolution compared to, \eg, the spectral decoding
method. We would like to point out that in contrast to the case of QMC, the
ED data to which the MEM is applied does not have a statistical error.

%
%
%
\section{Results}\label{sec:results}
%
%
%

It is long known that within the Holstein model~(\ref{eq:model:H}), a single
electron undergoes a cross over to a small polaron at a critical value of the
EP coupling.\cite{dRLa82} The important question to be addressed
here is what happens if the density of electrons (or polarons) is increased. 
We do not expect significant changes in the strong-coupling regime due to the
rather ``localized" nature of the small polaron state. On the contrary, for
intermediate EP interaction and $\omb<1$, each electron is
surrounded by a phonon cloud extending over several lattice sites.
Now, at larger densities, a substantial overlap of the single-particle
wavefunctions occurs, leading to a dissociation of the individual polarons
and finally to a restructuring of the whole many-particle ground
state. Furthermore, it has been shown that squeezing
effects reduce the polaronic band-narrowing in the many-electron
case,\cite{FeIhLoTrBu94} thereby favoring such a scenario.
Since within the Holstein model, a large polaron state is expected to be
stable in 1D, we will only consider this case in the sequel. 

In view of the situation in most polaronic materials, we mainly
consider the adiabatic regime, taking $\omb=0.4$. To illustrate the important
differences between $\omb\ll1$ and $\omb\gg1$, however, some results for the
antiadiabatic value $\omb=4$ will also be shown. 

\subsection{Limiting cases}\label{sec:results:limiting}

As we shall see below, the spectra in the intermediate coupling region turn
out to have a fairly complex structure. It is therefore helpful to begin with
the limiting cases of weak and strong EP coupling. 

\subsubsection{Weak coupling}

For weak coupling $\lambda=0.1$, the sign problem is not severe
(Sec.~\ref{sec:app:sign_results}) and the QMC simulations can easily be
performed for large lattices with $N=32$, thereby making the dispersion of
quasiparticle features well visible.  As indicated in
Eq.~(\ref{eq:app:maxent}), the MEM inversion yields $A(k,\om-\mu)$, where
$\mu$ denotes the chemical potential. 

\begin{figure}[t]
  \includegraphics[width=0.2385\textwidth]{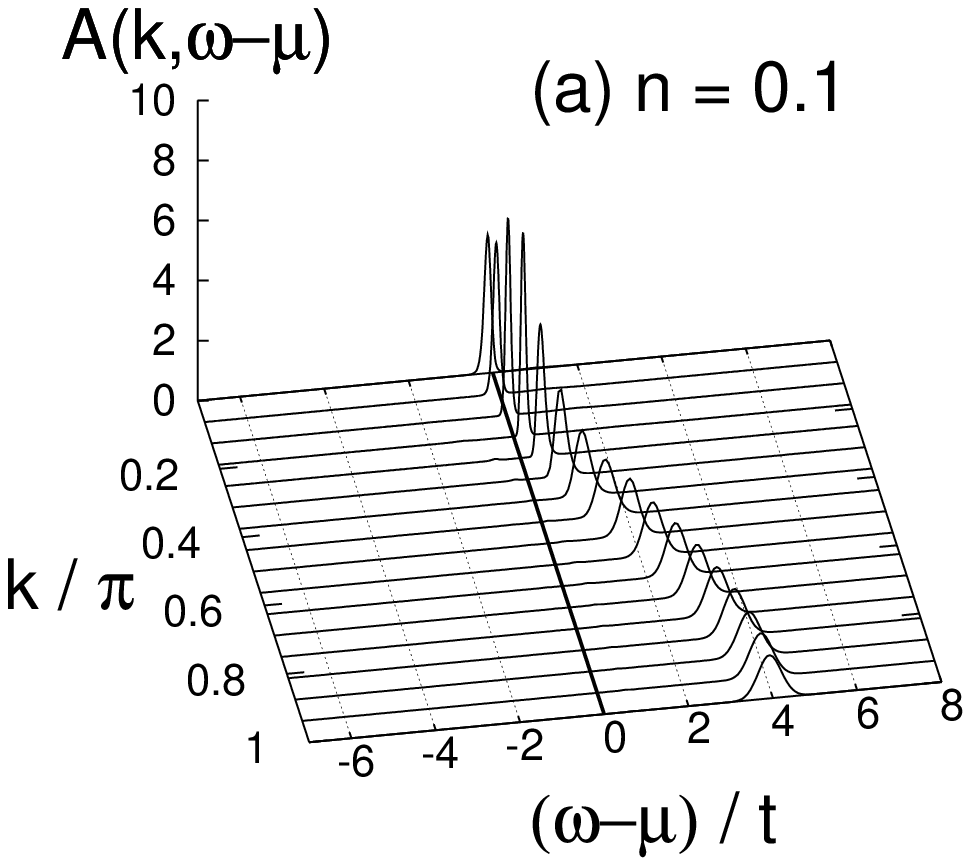}
  \includegraphics[width=0.2385\textwidth]{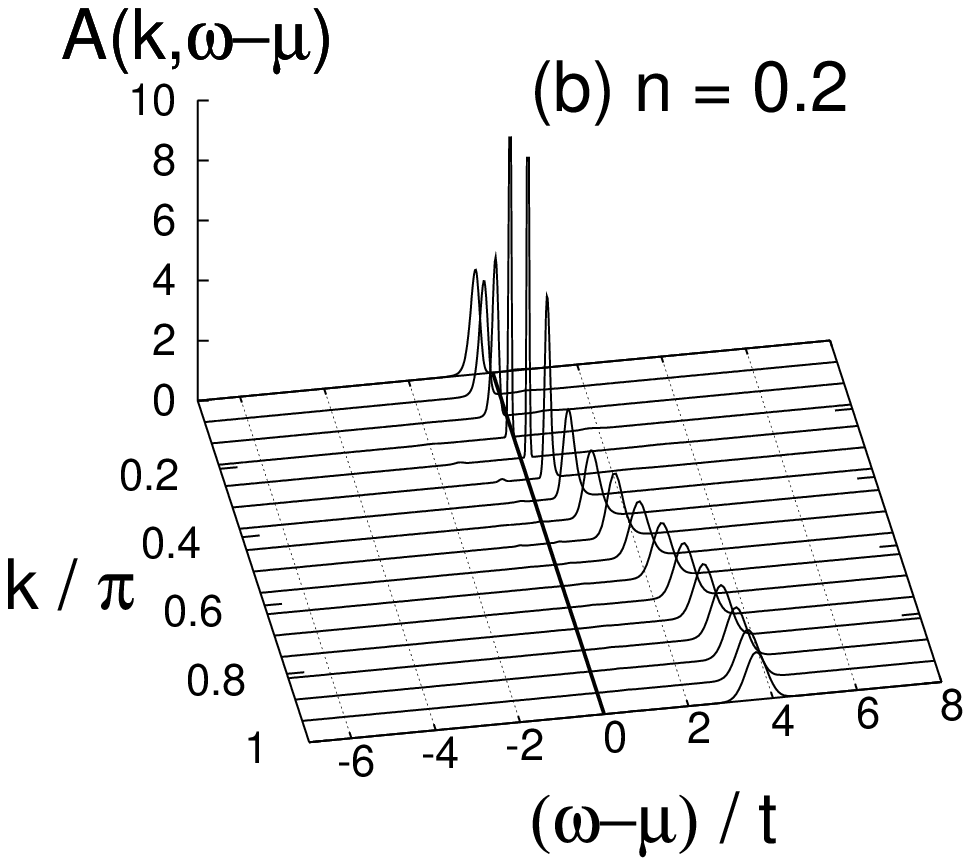}\\
  \includegraphics[width=0.2385\textwidth]{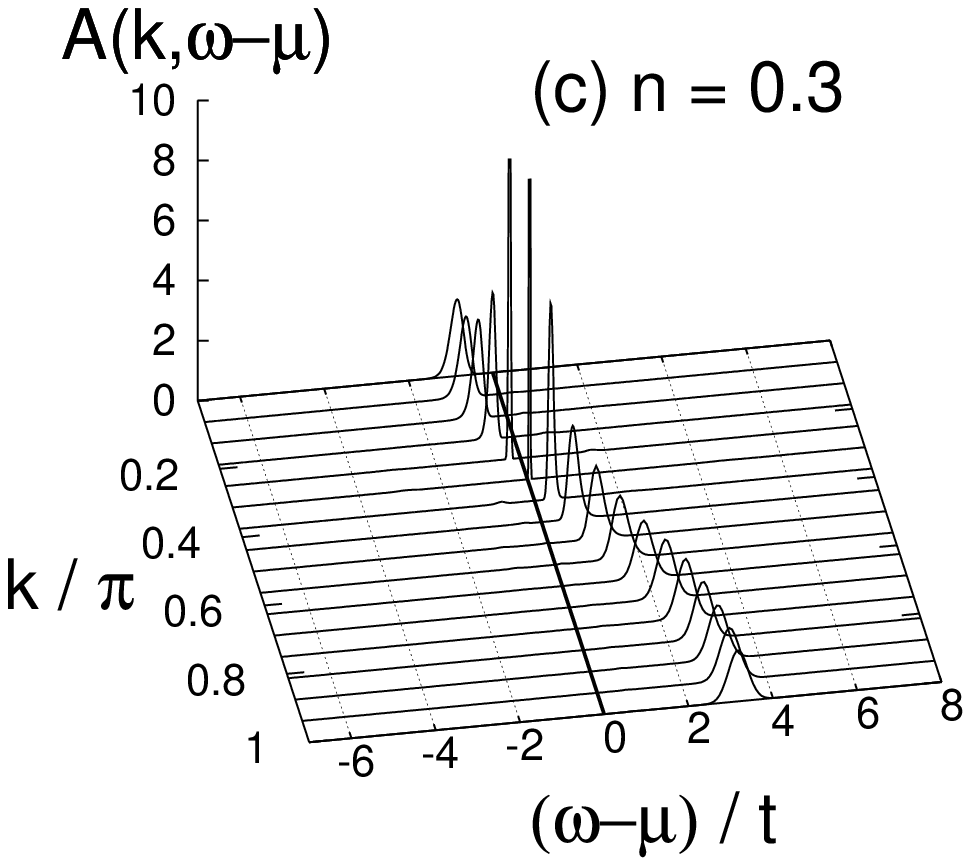}
  \includegraphics[width=0.2385\textwidth]{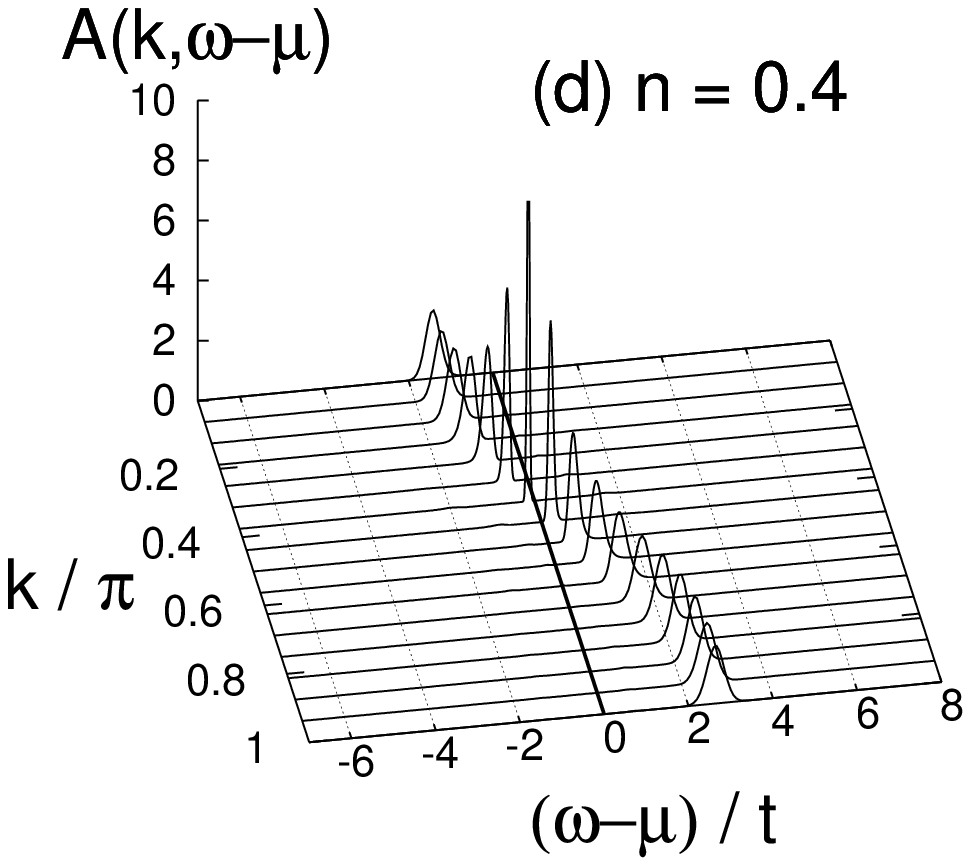}
  \caption{\label{fig:results:QMC_lambda0.1}
    One-electron spectral function $A(k,\om-\mu)$ from QMC for different band
    fillings $n$, $N=32$, $\beta t=8$, $\omb=0.4$, and weak coupling $\lambda=0.1$.} 
\end{figure}

Figure~\ref{fig:results:QMC_lambda0.1} shows the evolution of the
one-electron spectral function $A(k,\om-\mu)$ with increasing electron density
$n$. At first sight, we see that the spectra bear a close resemblance to the
free-electron case, \ie, there is a strongly dispersive band running from
$-2t$ to $2t$. The latter can be attributed to weakly dressed
electrons with an effective mass similar to the noninteracting case.  As
expected, the height (width) of the peaks increases (decreases) significantly
in the vicinity of the Fermi momentum $k_\text{F}$, which is determined by
the crossing of the band with the chemical potential. 

\begin{figure}[t]
  \centering
  \includegraphics[width=0.475\textwidth]{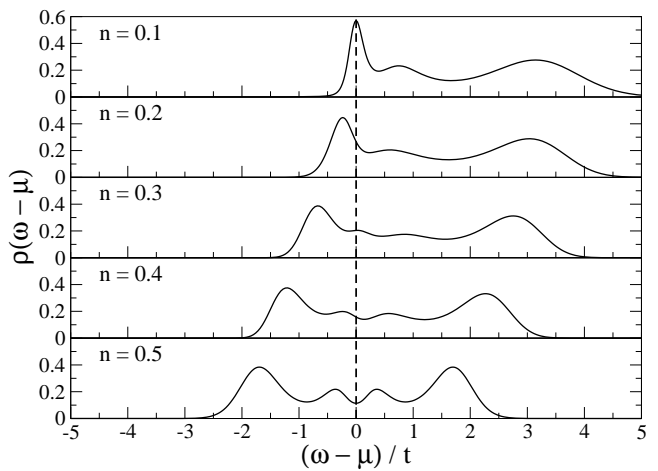}
\caption{\label{fig:results:dos_wc}
  One-electron density of states $\rho(\om-\mu)$ from QMC for different band
  fillings $n$, $N=32$, $\beta t=8$, $\omb=0.4$ and $\lambda=0.1$.} 
\end{figure}

Since the peaks in Fig.~\ref{fig:results:QMC_lambda0.1} are generally very
sharp, it is important to check the results from the inversion with the MEM. 
By computing the approximate number of electrons corresponding to the
fillings $n=0.1$\,--\,0.4 of the 32-site system, we can roughly determine
$k_\text{F}$.  Although the nonzero EP interaction will affect the band
dispersion $E(k)$, the weak coupling considered here justifies the use of a
rigid band approximation, as reflected in the free-particle character of the
spectra in Fig.~\ref{fig:results:QMC_lambda0.1}. 

For $n=0.1$, \ie, about three electrons for $N=32$, the Fermi momentum is
expected to lie between $k=\pi/16$ and $k=\pi/8$, corresponding to the second
and third curve from the top in Fig.~\ref{fig:results:QMC_lambda0.1}(a). In
fact, we see that the largest peaks are those for $k=\pi/16$ and $\pi/8$,
having almost the same height. Above and below these momenta, the peaks are
noticeably smaller.  A similar analysis carried out for the other fillings in
Figs.~\ref{fig:results:QMC_lambda0.1}(b)\,--\,(d) also shows a very good
agreement between the location of the highest peaks and the estimated values
of $k_\text{F}$. 

Finally, we would like to point out that the apparent absence of any phonon
signatures in Fig.~\ref{fig:results:QMC_lambda0.1} is not a defect of the
MEM, but results form the large scale of the $z$-axis chosen. As a
consequence, the peaks running close to the bare band dominate the spectra
and suppress any small phonon peaks present. If we enlarge the resolution,
for all densities $n=0.1$\,--\,0.4, we observe the band
flattening\cite{Stephan,WeFe97,HoAivdL03} at large wavevectors which
originates from the intersection of the approximately free-electron
dispersion with the optical phonon energy occurring at $\om-\mu=\om_0$. 

To complete our discussion of the weak-coupling regime, we show in
Fig.~\ref{fig:results:dos_wc} the one-electron density of states (DOS)
$\rho(\om-\mu)$ given by Eq.~(\ref{eq:app:dosgen}). Clearly, for small $n$,
there is a peak with large spectral weight at the Fermi level. In contrast,
for large $n$, the tendency toward formation of a Peierls-- (band--)
insulating state at $n=0.5$ suppresses the DOS at the Fermi
level, although we are well below the critical value of $\lambda$ at which
the cross over to the insulating state takes place at $T=0$.\cite{BuMKHa98}
The additional small features
separated from $\mu$ by the bare phonon energy $\om_0$ will be discussed in
the intermediate coupling case below.

\subsubsection{Strong coupling}
\begin{figure}[t]
  \includegraphics[width=0.2385\textwidth]{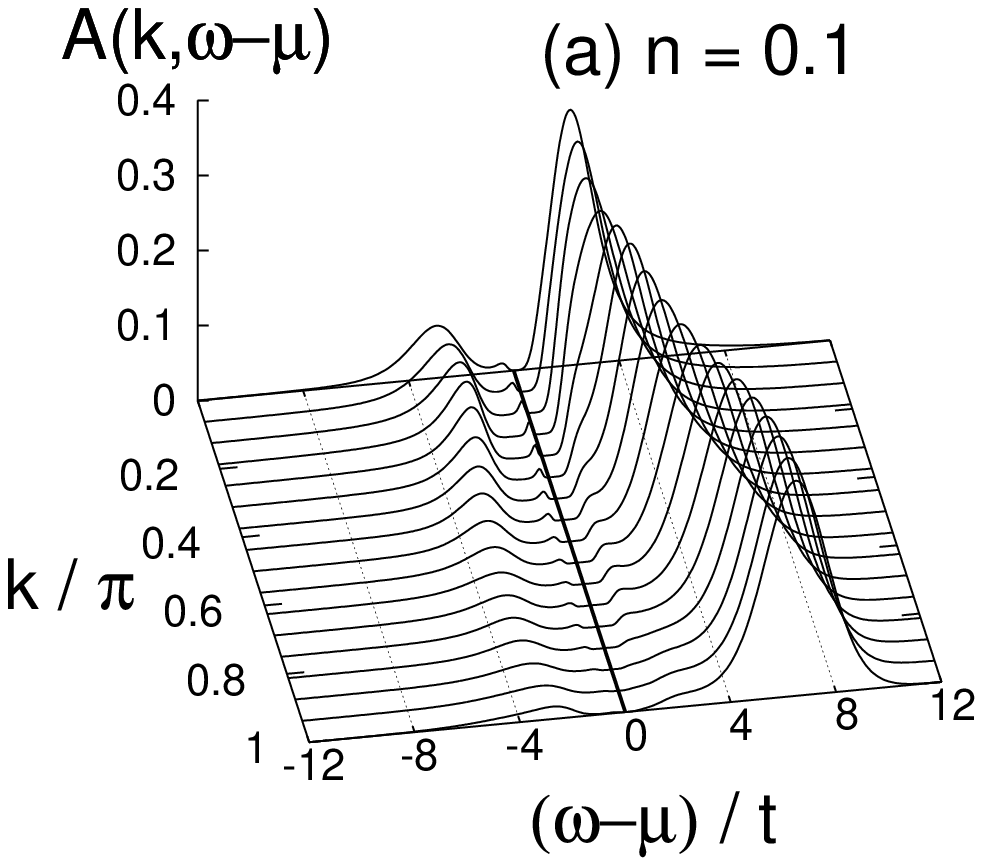}
  \includegraphics[width=0.2385\textwidth]{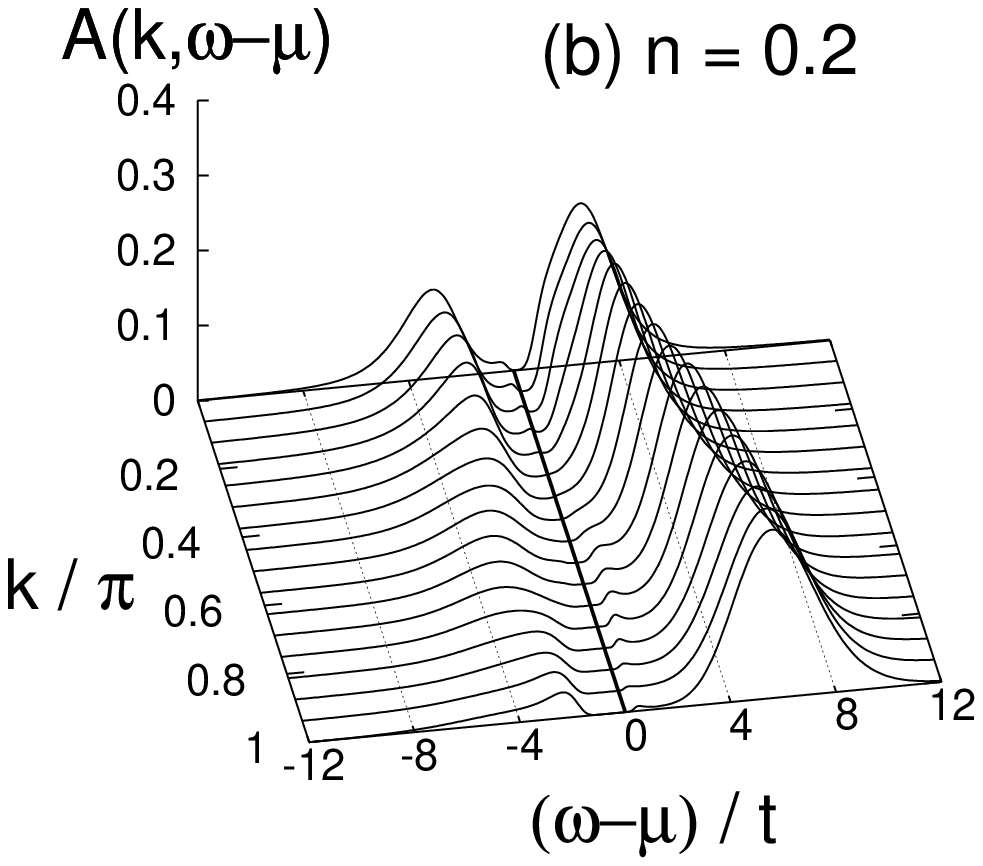}\\
  \includegraphics[width=0.2385\textwidth]{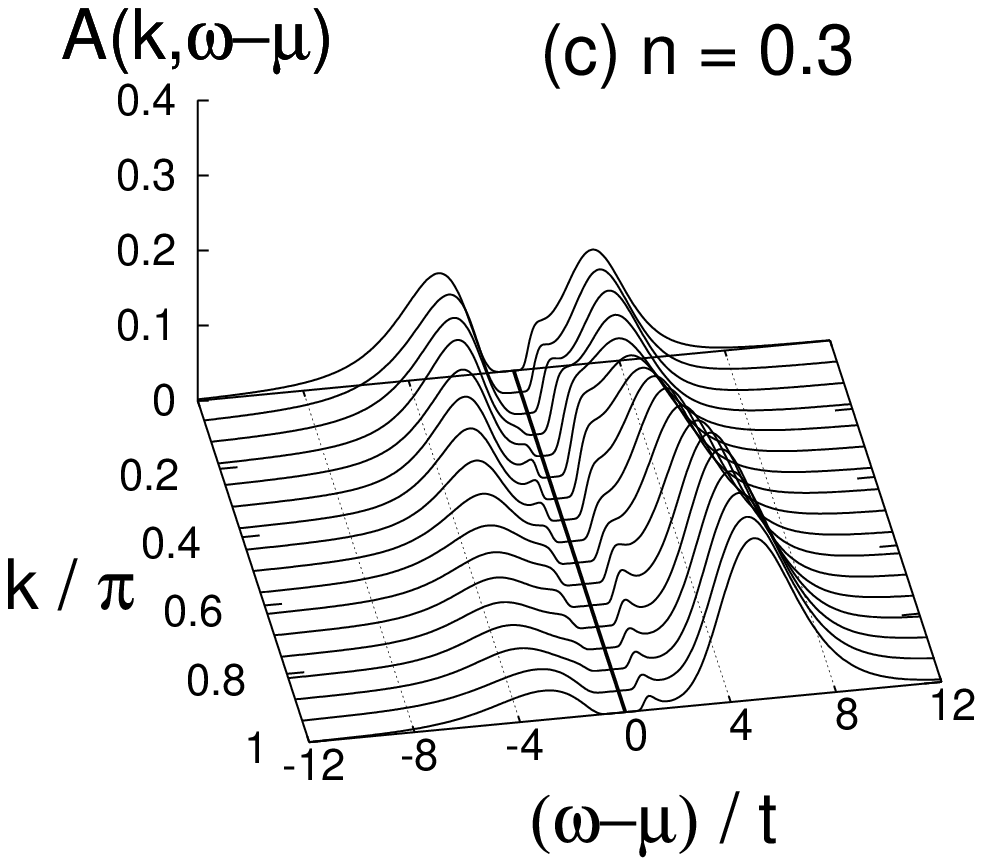}
  \includegraphics[width=0.2385\textwidth]{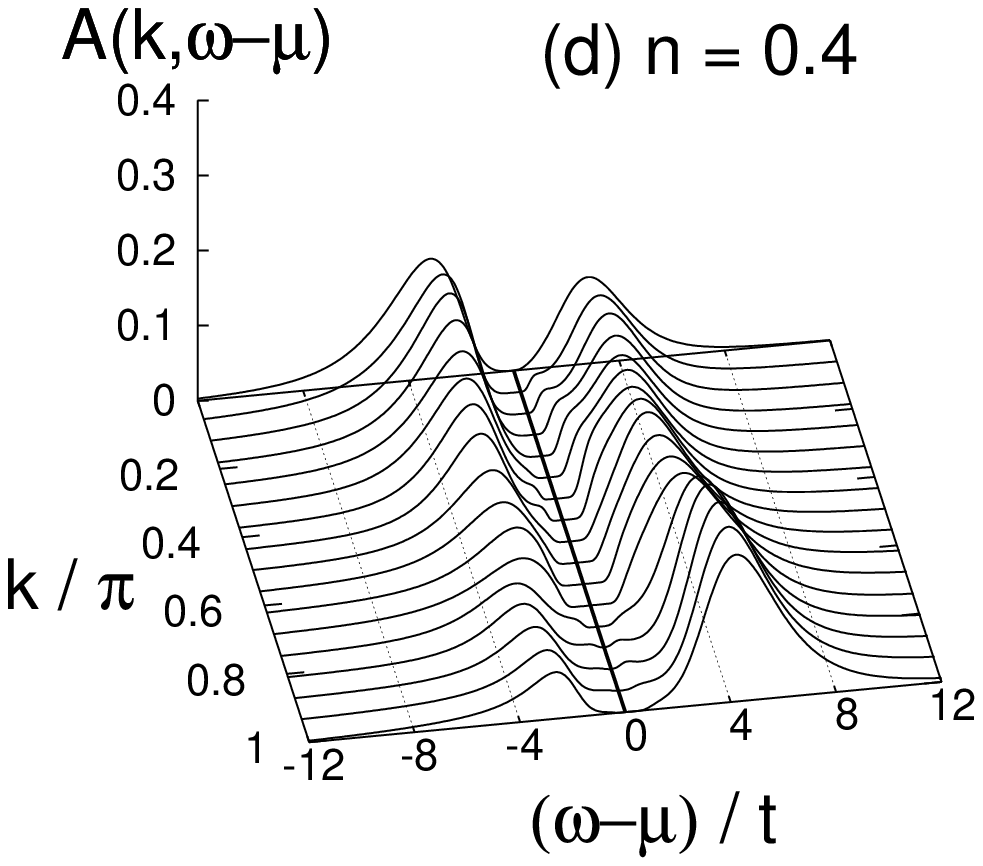}
  \caption{\label{fig:results:QMC_lambda2.0}
    One-electron spectral function $A(k,\om-\mu)$ from QMC for different band
    fillings $n$, $N=32$, $\beta t=8$, $\omb=0.4$, and strong coupling $\lambda=2.0$.} 
\end{figure}

We now turn our attention to the opposite limit of the spinless Holstein
model, namely the case of strong EP coupling $\lambda=2$. While calculations
with many other numerical methods such as, \eg, standard grand-canonical QMC
or ED, become very difficult in this regime, the QMC approach used here
yields quite accurate results even for large systems. The improved statistics
in the strong-coupling regime is also reflected in the dependence of the
average sign on $\lambda$ (cf. the discussion of
Fig.~\ref{fig:app:sign_beta_omega} in the Appendix).  As before, we study the
dependence of the spectral function on the band filling, again for $N=32$. 
The results are depicted in Fig.~\ref{fig:results:QMC_lambda2.0}. 

\begin{figure}[t]
  \includegraphics[width=0.2385\textwidth]{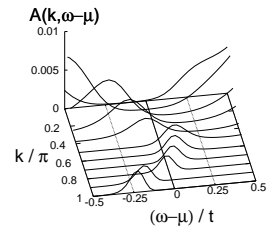}
  \caption{\label{fig:results:QMC_lambda2.0_polaron}
    Dispersion of the polaron band in $A(k,\om-\mu)$ for $n=0.05$, $N=20$ and
    $\beta t=10$ (drawn to a larger scale as compared to Fig. 3). All other
    parameters as in Fig.~\ref{fig:results:QMC_lambda2.0}.}
\end{figure}

In the low-density regime $n=0.1$ [Fig.~\ref{fig:results:QMC_lambda2.0}(a)],
we expect the well-known almost flat polaron band having exponentially
reduced spectral weight (given by $e^{-g^2}$ in the single-electron,
strong-coupling limit) which, nevertheless, can give rise to
coherent transport at $T=0$. Unfortunately, the QMC/MEM has difficulties
resolving such extremely weak signatures, especially at larger $n$, where the
statistical errors are more noticeable, and for temperatures such that the
thermal energy is comparable to the polaron band width. 
\footnote{Owing to the small weight contained in the polaron band, the
resulting band filling $n$ obtained by integration over the spectrum is not
sensitive enough to yield the correct value of $\mu$.} 
Therefore the polaron
band can hardly be resolved in Figs.~\ref{fig:results:QMC_lambda2.0}(a) or (b), and is 
not even detectable in Figs.~\ref{fig:results:QMC_lambda2.0}(c) and (d).
However, if we consider a density $n=0.05$, corresponding to a single
electron on a cluster with $N=20$, at a lower temperature $\beta t=10$,
the QMC results do indeed give a polaron band around $\mu$
(Fig.~\ref{fig:results:QMC_lambda2.0_polaron}). Due to the sign problem,
similar calculations for $n>0.1$ would be very demanding.

Besides, the spectrum consists of two incoherent
features located above and below the chemical potential, which reflect the phonon-mediated
transitions to high-energy electron states. Here, the maximum of the
photoemission spectra ($\om-\mu>0$) follows a tight-binding cosine dispersion. 
The incoherent part of the spectra is broadened according to the phonon
distributions. We would like to mention that the spectrum of a
single polaron at strong EP coupling has been studied analytically by
Alexandrov and Ranninger.\cite{AlRa92}

With increasing band filling [see panels (b) and (c)], the chemical potential
is shifted to lower energies $\propto \Ep n$, but is expected to be still
located in a narrow band with little spectral weight. Regardless of the
aforementioned  shortcomings, the QMC results provide a clear picture of the
physics. There exists a finite gap to the 
photoemission (inverse photoemission) parts of the spectrum, so that the system typifies as a
polaronic metal. We shall see below that a completely different behavior is
observed at intermediate coupling. Notice that the incoherent inverse photoemission
(photoemission) signatures are more pronounced at small (large) wavevectors.

Finally, for $n=0.4$ [Fig.~\ref{fig:results:QMC_lambda2.0}(d)], the
incoherent features lie rather close to the Fermi level, thus being
accessible by low-energy excitations. Now, the photoemission spectrum for
$k<\pi/2$ becomes almost symmetric to the inverse photoemission spectrum for
$k>\pi/2$ and already reveals the gapped structure which occurs at $n=0.5$
due to the CDW-formation accompanied by a Peierls distortion. 

\begin{figure}[t]
  \centering
  \includegraphics[width=0.475\textwidth]{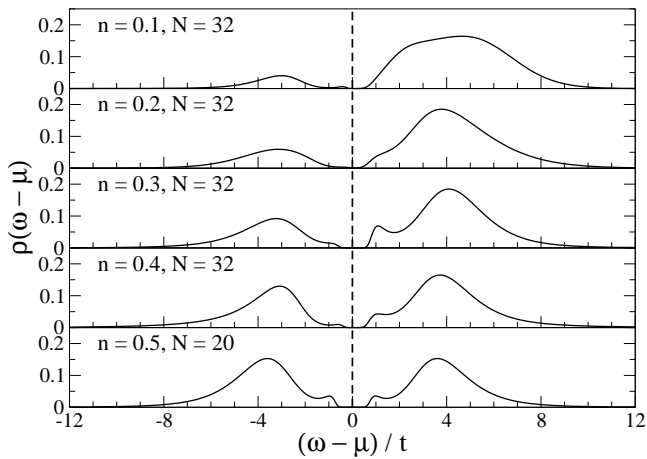}
\caption{\label{fig:results:dos_sc}
  One-electron density of states $\rho(\om-\mu)$ from QMC for different band
  fillings $n$ and cluster sizes $N$, $\beta t=8$, $\omb=0.4$ and $\lambda=2.0$.} 
\end{figure}

As in the weak-coupling case discussed above, the properties of the system
also manifest itself in the DOS, shown in
Fig.~\ref{fig:results:dos_sc}. Owing to the strong EP interaction,  the
spectral weight at the chemical potential is exponentially small for
all fillings $n$. 

At half filling, the DOS exhibits particle-hole symmetry, and the system can
be described as a Peierls insulator, consisting of a polaronic superlattice. 
In contrast to the weak-coupling case, the ground state is characterized as a
polaronic insulator rather than a band insulator.  Moreover, here the gap at
the Fermi level persists even at the finite temperature used in the
simulation, while for $\lambda=0.1$ and $n=0.5$ the insulating state found at
$T=0$ is destroyed by thermal fluctuations (Fig.~\ref{fig:results:dos_wc}). Note that, owing to the sign
problem, the result for $n=0.5$ in Fig.~\ref{fig:results:dos_sc} is for a
reduced cluster size $N=20$. However, in the strong-coupling regime
considered here, finite-size effects are expected to be very small. 

\begin{figure*}[t]
  \includegraphics[width=0.45\textwidth]{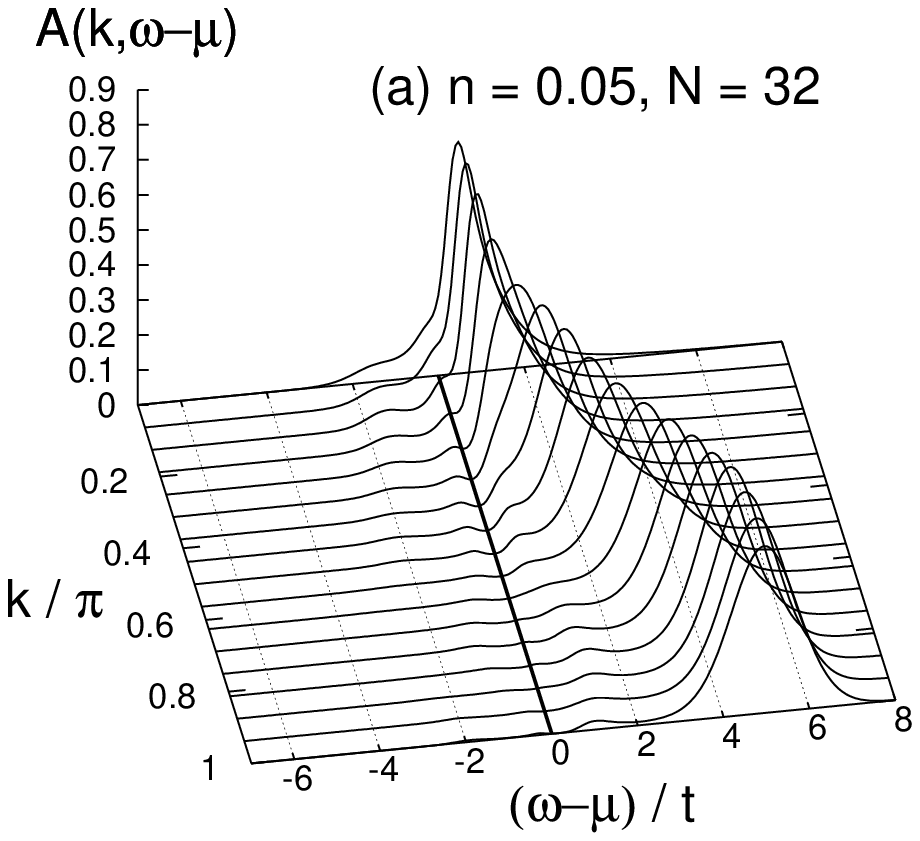}
  \includegraphics[width=0.45\textwidth]{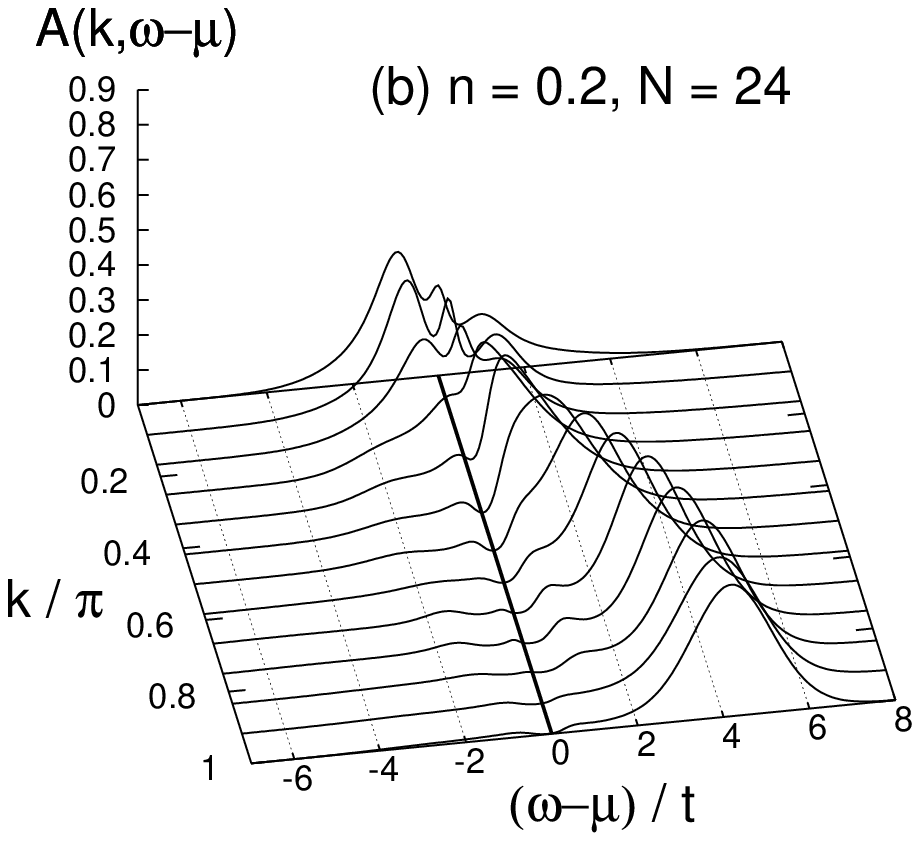}\\
  \includegraphics[width=0.45\textwidth]{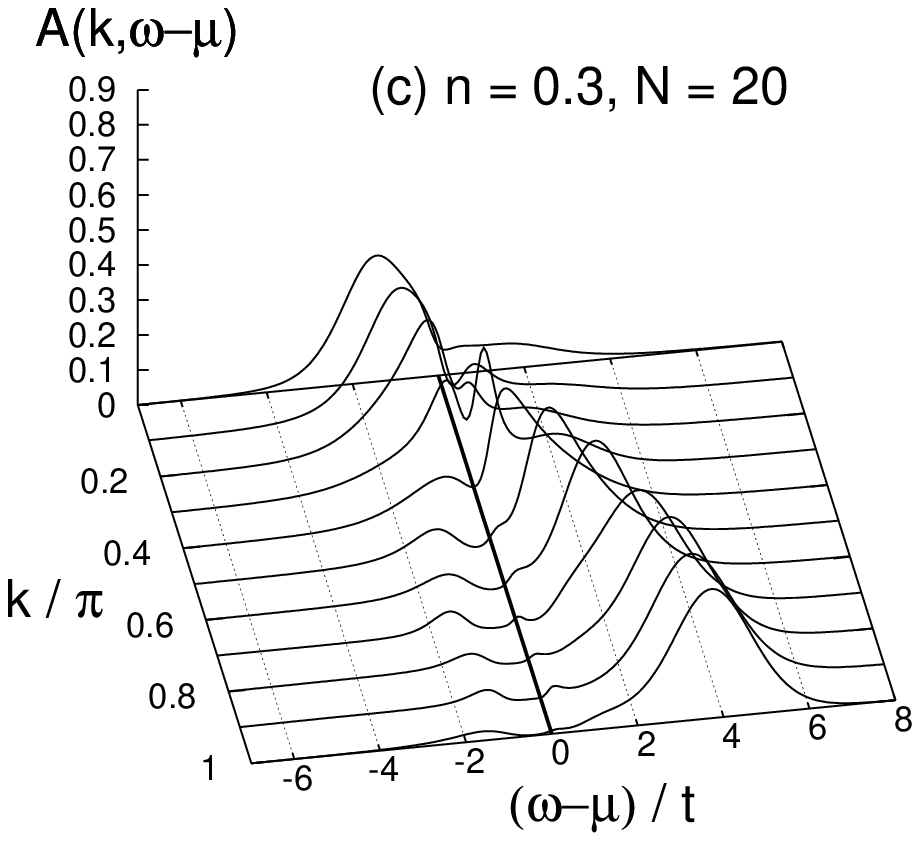}
  \includegraphics[width=0.45\textwidth]{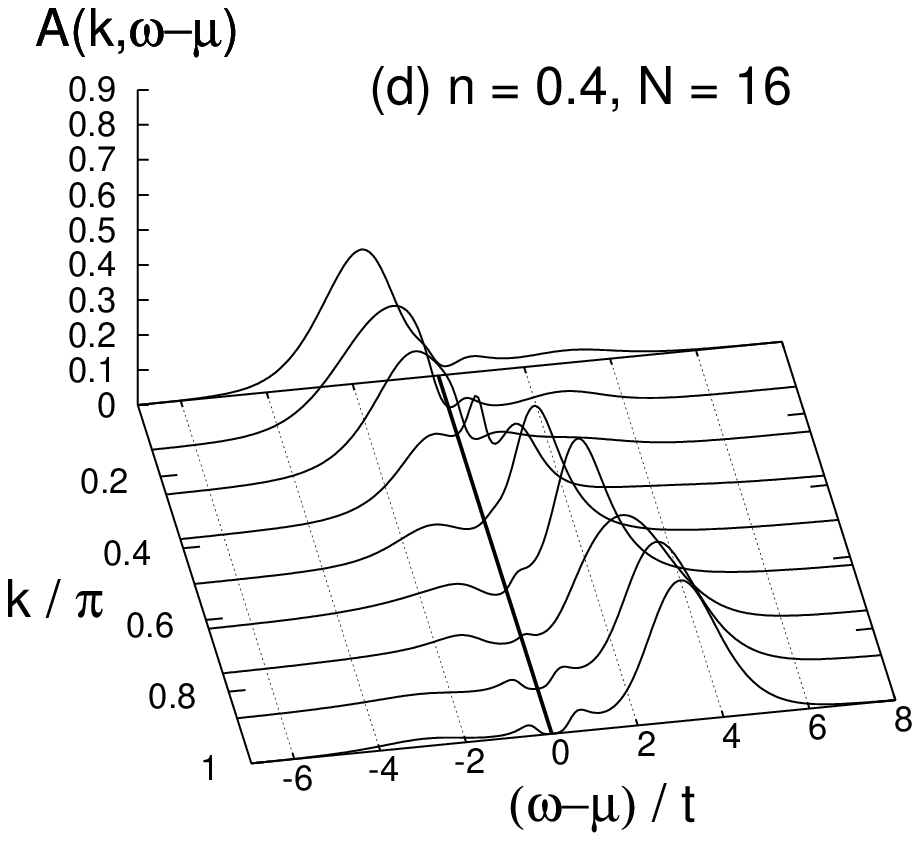}
  \caption{\label{fig:results:QMC_lambda1.0}
    One-electron spectral function $A(k,\om-\mu)$ from QMC for different band
    fillings $n$ and cluster sizes $N$, $\beta t=8$, $\omb=0.4$, and
    intermediate coupling $\lambda=1.0$.} 
\end{figure*}
%

\subsection{Intermediate coupling}\label{sec:results:intermediate}

As discussed in the introduction, a cross over from a polaronic state to a
system with weakly dressed electrons can be expected in the intermediate
coupling regime. Here we choose $\lambda=1$, which corresponds to the
critical value for the small-polaron cross over in the one-electron problem
(cf., \eg, Fig.~5 in Ref.~\onlinecite{HoEvvdL03}). 

\subsubsection{Quantum Monte Carlo}

We first discuss the QMC results. Owing to the sign problem, which is
particularly noticeable for $\lambda=1$ (see Appendix), we have to decrease
the system size as we increase the electron density $n$. 

Figure~\ref{fig:results:QMC_lambda1.0} shows the spectral function for
$\lambda=1$ and increasing band filling.  Owing to the overlap of large
polarons in the intermediate coupling regime, we start with the very low
density case $n=0.05$ [Fig.~\ref{fig:results:QMC_lambda1.0}(a)]. Compared to
the behavior for $\lambda=2$ [Fig.~\ref{fig:results:QMC_lambda2.0}(a)], we
notice that the polaron band now lies much closer to the incoherent features,
and that there is a mixing of these two parts of the spectrum at small values
of $k$. Nevertheless, the almost flat polaron band is well visible for large
$k$. 

With increasing density, the polaron band merges with the incoherent peaks at
higher energies. This is the above-anticipated density-driven cross over from a polaronic to a
(diffusive) metallic state, with the main band crossing the Fermi
level. At $n=0.4$, we find a band with large
effective width ranging from $\om\approx-4t$ to $\om\approx2t$.  A more
detailed discussion of the spectral function will be given below when we
present ED results. 

\begin{figure}[t]
  \centering
  \includegraphics[width=0.475\textwidth]{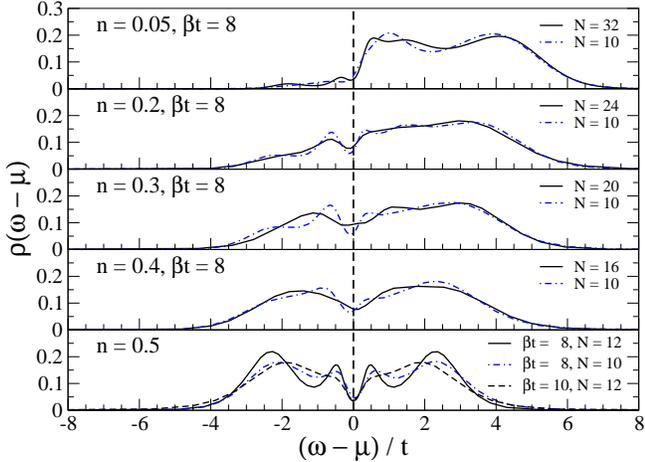}
\caption{\label{fig:results:dos_ic}(color online)
  One-electron density of states $\rho(\om-\mu)$ from QMC for different band
  fillings $n$, cluster sizes $N$ and inverse temperatures $\beta$. Here
  $\omb=0.4$ and $\lambda=1.0$.} 
\end{figure}

Further information about the density dependence can be obtained from the
one-electron DOS. The latter is presented in Fig.~\ref{fig:results:dos_ic}
for different fillings $n=0.05$\,--\,0.5.  As in
Fig.~\ref{fig:results:QMC_lambda1.0}, the cluster size is reduced with
increasing $n$ in order to cope with the sign problem. To illustrate the
rather small influence of finite-size effects, Fig.~\ref{fig:results:dos_ic}
also contains results for $N=10$. 

For low density $n=0.05$, the DOS in Fig.~\ref{fig:results:dos_ic} lies in
between the results for weak and strong coupling discussed above. Although
the spectral weight at the chemical potential is strongly reduced compared to
$\lambda=0.1$, $\rho(0)$ is still significantly larger than for $\lambda=2.0$. 

When the density is increased to $n=0.2$, the DOS at the chemical potential
increases, as a result of the dissociation of the polarons. As we
increase $n$ further, a pseudogap begins to form at $\mu$, which is a
precursor of the CDW gap at half filling and zero temperature. 

Finally, in the case of half filling $n=0.5$, the DOS has become symmetric
with respect to $\mu$.  There are broad features located either side of the
chemical potential, which take on maxima close to $\om-\mu=\pm\Ep$. However,
apart from the strong-coupling case, where the single-polaron binding energy
is still a relevant energy scale, the position of these peaks is rather
determined by the energy of the upper and lower bands, split by the
formation of a Peierls state. The gap of size $\sim\lambda$ expected for the
insulating charge-ordered state at $T=0$ is partially filled in due to the
finite temperature considered here. 

Furthermore, we find additional, much smaller features roughly separated from
$\mu$ by the bare phonon frequency $\om_0$, whose height decreases with
decreasing temperature, as revealed by the
results for $\beta t=10$ (Fig.~\ref{fig:results:dos_ic}). These peaks---not
present at $T=0$ (see Ref.~\onlinecite{SyHuBeWeFe04})---arise from thermally
activated transitions to states with additional phonons excited, and are
also visible in Figs.~\ref{fig:results:dos_wc} and~\ref{fig:results:dos_sc}. 
While for weak coupling [$\Ep=0.2$, $\lambda=0.1$, Fig.~\ref{fig:results:dos_wc}],
the maximum of these
features is almost exactly
located at $|\om-\mu|=\om_0$, it moves to $|\om-\mu|\approx1.25\om_0$ for
intermediate coupling [$Ep=2$, $\lambda=1$, Fig.~\ref{fig:results:dos_ic}],
and finally to $|\om-\mu|\approx2.5\om_0$ for
strong coupling [$\Ep=4$, $\lambda=2$, Fig.~\ref{fig:results:dos_sc}]. Although the
exact positions of the peaks are subject to
uncertainties due to the MEM, this reflects the shift of the maximum in the phonon
distribution function with increasing coupling. The MEM yields an envelope of
the multiple peaks separated by $\om_0$.

\begin{figure*}[t]
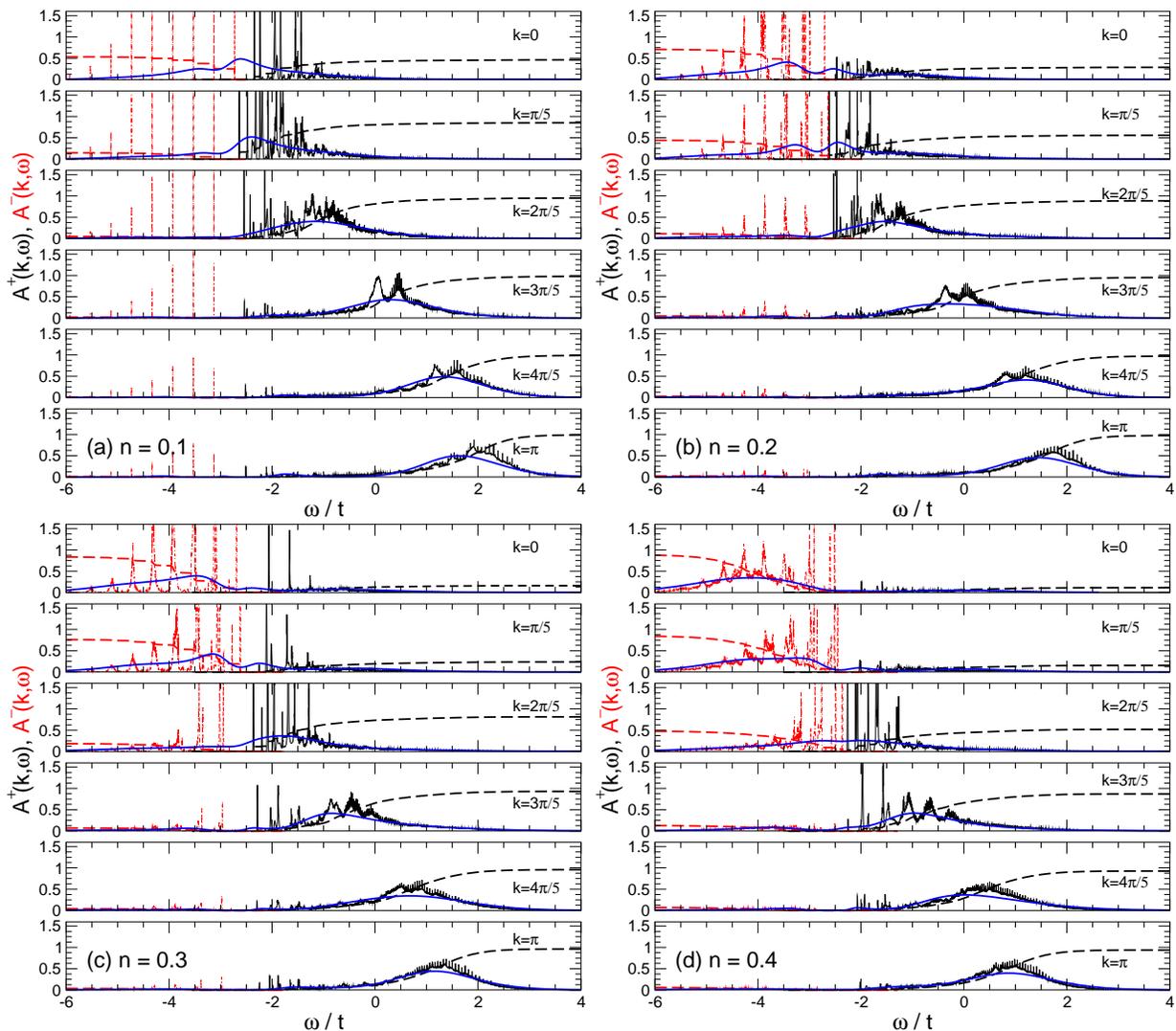

  \includegraphics[width=0.45\textwidth]{F8a.eps}
  \includegraphics[width=0.45\textwidth]{F8b.eps}\\
  \includegraphics[width=0.45\textwidth]{F8c.eps}
  \includegraphics[width=0.45\textwidth]{F8d.eps}
  \caption{\label{fig:results:ED_QMC_lambda1.0}(color online)
    One-electron spectral functions $A^-(k,\om)$ (red dot-dashed lines) and
    $A^+(k,\om)$ (black thin solid lines) from ED for different band fillings
    $n$, $N=10$, $\omb=0.4$, and intermediate coupling $\lambda=1.0$.  Also
    shown are the integrated spectral weights of $A^-$ and $A^+$ (dashed
    lines), as well as $A(k,\om)$ from QMC for $N=10$ and $\beta t=8$ (blue
    thick solid lines).} 
\end{figure*}

At this point, we would like to mention the exact relation\cite{Ko02}
\begin{equation}\label{eq:results:moment}
  M_1(k)
  =
  \int_{-\infty}^\infty\,d\om\, \om\, A(k,\om)
  =
  \ek - \mu - 2\Ep n
\end{equation}
for the first moment of the one-particle spectral function of the spinless
Holstein model, with $\ek=-2t\cos k$. While the zeroth moment is identical to
the normalization of $A(k,\om)$ for each $k$, $M_1(k)$ depends in a nontrivial way
on the parameters of the system. In principle, Eq.~(\ref{eq:results:moment})
may be used as a boundary condition in the MEM.\cite{Wh91} However, in the
present case, we have found that while the normalization is virtually exact
for all results shown, the first moment deviates from the exact values. 
Moreover, the use of the exact values for $M_1$ as a condition in the
inversion causes the MEM not to converge for some parameters. Additional
calculations on small systems reveal that this disagreement results from the
finite Trotter error, despite the relatively small value $\dtau=0.1$ used. 
Away from parameters such that $M_1\approx0$, $M_1$ as determined from the
MEM spectra deviates from the exact values by maximally 10\,--\,20 percent. 
Nevertheless, the peak positions and weights in the spectra shown are fairly
accurate, as illustrated by the comparison with ED below. Finally, we would
like to point out that calculations of $A(k,\om)$ for $\dtau<0.1$ become very
time-consuming due to the increasing requirement in disk space (see Appendix)
and longer simulation times. 

Despite the wealth of information contained in the QMC results presented so
far, it is obvious that an interpretation of the various excitations is far from
being easy. In addition to the finite temperature, the QMC method used here
does not allow to separately calculate the photoemission and inverse
photoemission parts $A^-$ and $A^+$, respectively. Moreover, the use of the
MEM limits the energy resolution and introduces uncertainties concerning the
positions and weights of structures in the spectra. These circumstances make it difficult,
\eg, to distinguish between phononic and electronic contributions, especially
if $\lambda\approx\omb$. To gain further insight, we therefore supplement the
QMC data with ED results.

\subsubsection{Exact diagonalization}

In order to overcome the abovementioned limitations of QMC, here we use ED in
combination with the KPM, as described in Sec.~\ref{sec:methods:ED}, to
calculate the one-particle spectral functions defined by
Eq.~(\ref{eq:methods:Apm}). To test the quality of the QMC spectra, we
directly compare the two methods, keeping in mind that the QMC calculations
have been performed at a finite temperature, whereas ED yields ground-state
results. 

\begin{figure}[t]
  \includegraphics[width=0.45\textwidth]{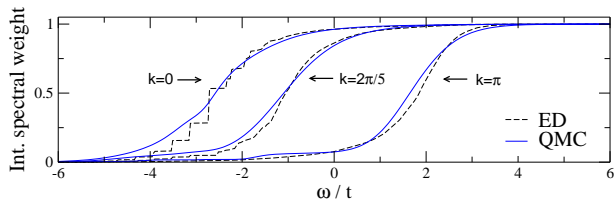}
  \caption{\label{fig:results:ED_QMC_weight}(color online)
  Comparison of the integrated spectral weight obtained from the QMC and ED
  data of Fig.~\ref{fig:results:ED_QMC_lambda1.0}(a) for three different momenta
  $k$.} 
\end{figure}

Figure~\ref{fig:results:ED_QMC_lambda1.0} displays the ED results for the
same parameters as in Fig.~\ref{fig:results:QMC_lambda1.0}, \ie, $\omb=0.4$,
$\lambda=1$, and $n=0.1$\,--\,0.4.  Additionally shown are the integrated
spectral weights of $A^-$ and $A^+$ (dashed lines), as well as QMC results
for the same system size and $\beta t = 8$ (solid thick lines), which have
been shifted by $\mu$. The Fermi level lies between the highest-energy peak
in $A^-$ and the lowest-energy peak in $A^+$. 

We see from Fig.~\ref{fig:results:ED_QMC_lambda1.0} that the QMC results are
compatible with the ED data. Because of the finite temperature and the use
of the MEM, QMC, of course, cannot reproduce the sharp ED peaks. To further
test the agreement between the two methods, we compare in
Fig.~\ref{fig:results:ED_QMC_weight} the integrated spectral weight of the
data shown in Fig.~\ref{fig:results:ED_QMC_lambda1.0}(a), and find a very
satisfactory accord. Moreover, this agreement becomes even better for larger
densities $n$, for which the sharp peaks in the ED data evolve into bands
[see Fig.~\ref{fig:results:ED_QMC_lambda1.0}(b)\,--\,(d)].

In addition to the possibility of distinguishing between $A^+$ and
$A^-$---crucial for an identification of polaron bands at intermediate
coupling---the higher energy resolution of ED also allows to resolve
signatures of phonon excitations. 

Starting with the case $n=0.1$ [Fig.~\ref{fig:results:ED_QMC_lambda1.0}(a)],
we notice that there is a polaron band at the Fermi level, excitations
to/from which are given by the highest (lowest) peak of $A^-$ ($A^+$). The
photoemission part $A^-$ reflects the Poisson-like
phonon distribution of the one-electron ground state ($n=0.1$ and $N=10$). 
The integrated spectral weight gives a measure for the electronic weight of
the various poles in the one-electron spectrum. For example, it reveals that
the phonon peaks in $A^-$ have very little weight. In fact, the integrated
weight jumps to a finite value at the first peaks near the Fermi level for
$k=0$ and $k=\pi/5$, while it changes very little as one moves further down
in energy. For $k>\pi/5$, the small spectral weight contained in $A^-$ is
continuously distributed among the phonon peaks. 

We also observe the well-known flattening of the polaron band at large values
of $k$.  Similar to the one-electron case,\cite{HoAivdL03} the low-energy
states have mostly electronic character at small $k$, and become mostly
phononic at large values of $k$.  While this effect is expected to occur in
the low-density regime, we find that is persists even for $n=0.4$
[Fig.~\ref{fig:results:ED_QMC_lambda1.0}(d)]. Moreover,
Fig.~\ref{fig:results:ED_QMC_lambda1.0} reveals that the maximum in the
incoherent contribution follows closely the free-electron dispersion for all
densities $n=0.1$\,--\,0.4. 

With increasing electron density, the equally spaced peaks in $A^-$ broaden
significantly, until at $n=0.4$ they have merged to form a broad band. The
polaron band is still visible at $n=0.2$
[Fig.~\ref{fig:results:ED_QMC_lambda1.0}(b)], but becomes indistinguishable
from the incoherent excitations at even larger $n$.  Eventually, at $n=0.4$
[Fig.~\ref{fig:results:ED_QMC_lambda1.0}(d)], we find an almost symmetric
behavior of $A^+$ and $A^-$ with respect to the Fermi level at $k=2\pi/5$. 
Compared to the strong-coupling case
[Fig.~\ref{fig:results:QMC_lambda2.0}(d)], there exist incoherent
excitations with almost zero energy.  This indicates the metallic character
of the system expected for large polaron densities and intermediate
EP coupling. 

\subsection{Antiadiabatic regime}\label{sec:results:nonadiabatic}

%
\begin{figure}[t]
  \includegraphics[width=0.45\textwidth]{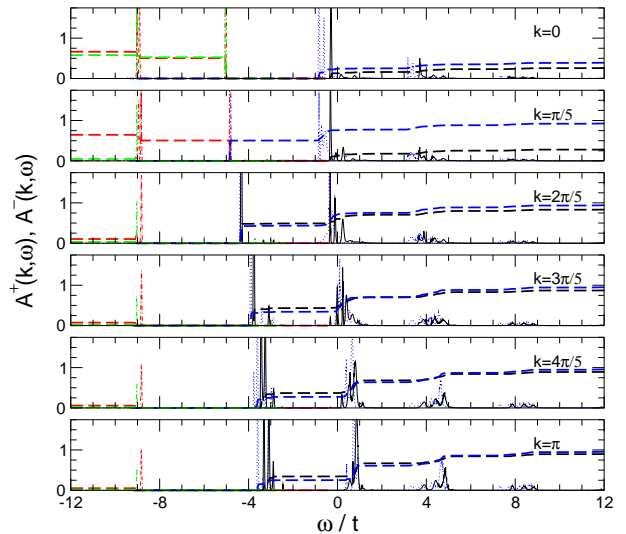}
  \caption{\label{fig:results:w4}(color online)
    One-electron spectral functions $A^-(k,\om)$ and $A^+(k,\om)$ from ED for $N=10$, 
    $\omb=4.0$, and  $\lambda=2.0$. Here $n=0.1$ ($A^-$: red dot-dashed lines,
    $A^+$: black solid lines) and $n=0.3$ ($A^-$: green dashed two-dotted lines, $A^+$:
    blue dotted lines). Dashed lines correspond to the integrated spectral weight
    of $A^-(k,\om)$ and $A^+(k,\om)$, respectively.} 
\end{figure}

Until here, we have only presented results for the adiabatic case $\omb=0.4$. 
Although the latter is most relevant for strongly correlated materials, it is
important to compare the above findings to the antiadiabatic strong-coupling
regime.  To this end, we chose $\omb=4$ as well as $\lambda=2$. The
condition for the formation of a single small polaron in this case,
however, is $g\gtrsim1$, because the phonon frequency is the important energy scale. 
In contrast, the criterion $\lambda>1$ is purely based on the balance of kinetic
and potential energy.\cite{WeFe97,CaStGr97,CiPaFrFe97} 

The spectral function for $n=0.1$ and $n=0.3$ is displayed in
Fig.~\ref{fig:results:w4} for $g=1$, \ie, at the transition point where
self-trapping occurs for a single particle. Contrary to the case $\lambda=1$
and $\omb=0.4$ considered above, the absorption features are now
separated by the phonon energy $\om_0$, with their width being determined by
electronic excitations. 

A comparison of the spectral functions for $n=0.1$ and $n=0.3$ in
Fig.~\ref{fig:results:w4} indicates that there is no density-driven cross
over of the system as observed in the adiabatic case. In particular, owing to
the large phonon energy, there are no low-energy excitations close to the
polaron band, so that the latter remains well separated from the incoherent
features even for $n=0.3$. Furthermore, the spectral weight of the polaron
band also remains almost unchanged as we increase the density from $n=0.1$
to $n=0.3$. Consequently, almost independent small
polarons are formed also at finite electron densities, in accordance with
previous findings of Capone \etal\cite{CaGrSt99} for small systems. 

\subsection{Kinetic energy}\label{sec:results:Ek}

\begin{figure}[t]
  \centering
  \includegraphics[width=0.45\textwidth]{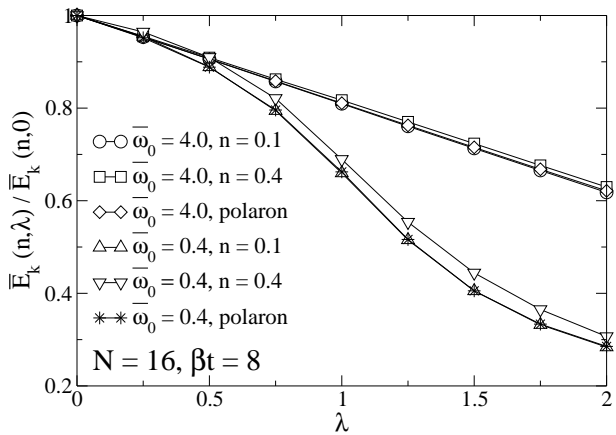}
\caption{\label{fig:results:Ek}
  Normalized kinetic energy as a function of EP coupling
  $\lambda$ for different band fillings $n$ and adiabatic ratios $\omb$. 
  Also shown are the results for a single polaron, obtained with the method
  of Ref.~\onlinecite{HoEvvdL03}. Errorbars are smaller
  than the symbols used and $\dtau=0.05$.} 
\end{figure}

Previously, the electronic kinetic energy has been studied extensively to
monitor the polaron and bipolaron cross over in the Holstein model with one
and two electrons (see, \eg, Refs.~\onlinecite{dRLa82}
and~\onlinecite{deRaLa86}). Here, at least in the
adiabatic regime, polaron formation manifests itself in
a drop of the kinetic energy around the critical coupling. In the following, however, we illustrate that the kinetic
energy [Eq.~(\ref{eq:app:Ek})] does not reflect the significant changes in
the one-particle spectrum we found with increasing density in the
intermediate coupling regime (see Sec.~\ref{sec:results:intermediate}). 

To compare different band fillings, we consider the normalized quantity
$\Ek(n,\lambda)/\Ek(n,0)$, shown in Fig.~\ref{fig:results:Ek} as a function
of EP coupling for different fillings and phonon frequencies,
for a fixed system size $N=16$ and $\beta t=8$. 

In the adiabatic regime $\omb=0.4$, owing to the overlap of the phonon clouds
surrounding the individual quasiparticles, the carriers clearly become more
mobile with increasing density, but the change in the kinetic energy is
nevertheless surprisingly small. In addition to these two fillings, we have
also included in Fig.~\ref{fig:results:Ek} the result for a single polaron
with the same parameters, calculated with the QMC method of
Ref.~\onlinecite{HoEvvdL03}, which is indistinguishable from the result for
$n=0.1$. 

As expected from the almost identical spectra for $n=0.1$ and 0.3 presented
in Sec.~\ref{sec:results:nonadiabatic} for the antiadiabatic regime, $\Ek$ is
almost independent of $n$ in this case.

%
%
%
\section{\label{sec:conclusions}Conclusions}
%
%
%

Strongly correlated electron-phonon systems in dependence on the carrier
density have been studied by means of unbiased numerical methods. Using
exact diagonalization together with the kernel polynomial method, as well as
a novel quantum Monte Carlo approach, we were able to present accurate
results for the one-dimensional spinless Holstein model in the most difficult
regime of small phonon frequencies and intermediate to strong electron-phonon
interaction. 

We have calculated the one-particle spectral function, the density of states
and the electronic kinetic energy at different band fillings. While exact
diagonalization is restricted to relatively small clusters, thereby limiting
the momentum resolution of the spectra, quantum Monte Carlo can be used to
study larger systems, but yields a lower energy resolution for dynamical
quantities. The reliability of the results from the maximum entropy inversion
has been scrutinized, and we find a very good agreement between the two
methods. 

In the adiabatic case, characteristic for, \eg, the manganites, we observe a
carrier density-driven cross over from a polaronic state to a metallic system
at weak and intermediate electron-phonon coupling.
Note that the nature of the quasiparticles is changed in the course of this
transition. This points toward an
insufficiency of simple single-polaron theories to explain experimental
results for polaronic materials. In fact, very similar effects due to carrier
interaction have recently been observed experimentally for the
manganites.\cite{HaMaDeLoKo04,HaMaLoKo04}

On the contrary, for large phonon frequencies and strong coupling, the
individual polarons remain virtually unaffected upon increasing the electron
density.  Consequently, the system remains polaronic even at large band
fillings. Finally, the kinetic energy is shown to be too insensitive in order
to reveal the above cross over. 

For weak electron-phonon coupling, the system becomes a Peierls or band
insulator at half filling, whereas for strong coupling a polaronic insulating
state with an exponentially small polaron band at the Fermi level, and a
finite gap to the incoherent excitations, prevails for all fillings. 

Finally, reliable calculations of the optical conductivity are highly
desirable in order to explain the large amount of available experimental data
for polaronic materials. Work along this line is in progress. 

\begin{acknowledgments}
  
  This work was supported by the Austrian Science Fund (FWF), project
  No.~P15834, the Deutsche Forschungsgemeinschaft through SPP1073, and the
  Bavarian Competence Network for High Performance Computing
  (KONWIHR). M.~H. is grateful to DOC (Doctoral Scholarship Program of the Austrian Academy of
  Sciences), and to HPC-Europa. H.~F. acknowledges the hospitality at Graz
  University of Technology. We would like to thank A. Alvermann, H. G.
  Evertz, T. Lang and A. Wei{\ss}e for useful discussion. Furthermore, we
  acknowledge generous computer time granted by the LRZ Munich.

\end{acknowledgments}

\appendix

%
%
\section{Quantum Monte Carlo}\label{sec:app:QMC}
%
%
%

In this Appendix we extend the one-electron QMC algorithm developed in
Ref.~\onlinecite{HoEvvdL03}, henceforth also referred to as I, to the
spinless Holstein model with many electrons. We begin by applying the
Lang-Firsov transformation\cite{LangFirsov} $\hat{\nu}=\exp(\rmi\gamma\sum_j
\on_j\op_j)$ with $\gamma^2=4\lambda\rD/\omb$ to the
Hamiltonian~(\ref{eq:model:H}) in first quantization and with dimensionless
phonon variables $\ox$ and $\op$, which takes the form\cite{HoEvvdL03}
\begin{equation}\label{eq:app:H}
  H
  =
  -t\sum_{\las i,j\ras} c^\dag_i c^{\nag}_j
  +\frac{\om_0}{2}\sum_i (\op^2_i + \ox^2_i)
  -\alpha \sum_i \on_i \ox_i
  \,. 
\end{equation}
The coupling constant $\alpha$ is related to that in Eq.~(\ref{eq:model:H})
by $\alpha=\sqrt{2}g\om_0$. Using the result $\tilde{H}=\hat{\nu}H\hat{\nu}^\dag$
(see I), we define the grand-canonical Hamiltonian
\begin{eqnarray}\nonumber\label{eq:app:Hspinless}
  \tilde{\H}
  &=&
  \tilde{H} - \mu \sum_i \on_i
  \\\nonumber
  &=&
  \underbrace{%
  -t\sum_{\las i,j\ras} c^\dag_i c^{\nag}_j e^{\rmi\gamma(\op_i - \op_j)}
  }_{\tilde{K}}
  +
  \underbrace{%
  \frac{\om_0}{2}\sum_i (\op^2_i + \ox^2_i)}
  _{P \equiv P_p + P_x}
  \\
  &&-
  \underbrace{%
  (E_\text{P} + \mu) \sum_i \on_i
  }_{\tilde{I}}
  \,,
\end{eqnarray}
where $\mu$ denotes the chemical potential and $\Ep$ is the polaron binding
energy. 

For the case of a half-filled band, the chemical potential is given by
$\mu=-\Ep$. Away from $n=0.5$, $\mu$ has to be adjusted accordingly to yield
the right density of electrons $n$. 

We would like to point out that in the spinfull case, the transformed
Hamiltonian would contain an attractive Hubbard term. For the determinant QMC
method to be applicable, the latter has to be decoupled using auxiliary
fields.\cite{wvl1992} In contrast, for the spinless model considered here,
the algorithm is almost identical to the one-electron case.  The following
derivation assumes a hypercubic lattice in D dimensions, consisting of
$N^\rD$ sites. 

\subsection{Partition function}\label{sec:app:QMC_partition}

We use the Suzuki-Trotter decomposition\cite{wvl1992}
\begin{equation}\label{eq:app:suzuki-trotter}
  e^{-\beta\tilde{\H}}
  \approx
  \left(
    e^{-\dtau\tilde{K}}
    e^{-\dtau P_\text{p}}
    e^{-\dtau P_\text{x}}
    e^{-\dtau \tilde{I}}
  \right)^L
  \,,
\end{equation}
with $\dtau=\beta/L$. The trace appearing in
the partition function $\Z = \tr e^{-\beta\tilde{\H}}$ can be split up into a
bosonic and a fermionic component leading to 
\begin{eqnarray}\nonumber
  \Z_L
  =
  &&\tr_\text{f}\int\,d p_1d p_2\cdots d p_L
  \\\nonumber
  &&\times
  \prod_{i=1}^L
  \bra{p_i}
  e^{-\dtau\tilde{K}}
  e^{-\dtau P_\text{p}}
  e^{-\dtau P_\text{x}}
  e^{-\dtau \tilde{I}}
  \ket{p_{i+1}}
  \,,
\end{eqnarray}
where $d p_\tau\equiv\prod_i d p_{i,\tau}$, and with periodic boundary
conditions $p_{L+1}=p_1$. We have chosen the time ordering, which is
arbitrary, in increasing order. The phonon coordinates $\ox$ in
$\Z_L$ can be integrated out analytically in the same manner as in I. 
Moreover, the momenta $\op$ can be replaced by their eigenvalues on each time
slice, and the partition function takes the form
\begin{equation}
  \Z_L
  =
  C \int\,\D p\,e^{-\dtau S_\text{b}}
  \tr_\text{f}(\oB_1\oB_{2}\cdots\oB_L)
\end{equation}
with $\D p=\prod_\tau d p_\tau$, $C=[2\pi/(\om_0\dtau)]^{NL/2}$ and
\begin{equation}\nonumber
  \oB_\tau
  =
  e^{-\dtau\tilde{K}_{0,\tau}}
  e^{-\dtau\tilde{I}}
  \,,\quad
  \tilde{K}_{0,\tau}
  =
  -t\sum_{\las i,j\ras} c^\dag_i c^{\nag}_j e^{\rmi\gamma(p_{i,\tau} - p_{j,\tau})}
  \,. 
\end{equation}
The bosonic action $S_\text{b}$ may be expressed in terms of principal
components (see I). The fermion degrees of freedom can be
integrated out exactly leading to\cite{BlScSu81}
\begin{eqnarray}\label{eq:app:omega}\nonumber
   \tr_\text{f} (\oB_1 \cdots \oB_L)
   &=&
   \det (1 +  B_1\,\cdots\,B_L)
   \\
   &\equiv&
   \det (1 + \Omega)\,,
\end{eqnarray}
where the $N^\rD\times N^\rD$ matrix $B_\tau$ is given by
\begin{equation}\label{eq:app:matprod}
  B_\tau
  =
  D_\tau\,\kappa\,D^\dag_\tau\,\V
\end{equation}
with
\begin{eqnarray}\nonumber
  \kappa_{ij}
  =
  (e^{\dtau t h^\text{tb}})_{ij}
  &\,,\quad&
  (D_\tau)_{ij}
  =
  \delta_{ij}\,e^{\rmi\gamma p_{i,\tau}}
  \,,
  \\\nonumber
  \V_{ij}
  &=&
  \delta_{ij}\,e^{\dtau(E_\text{P}+\mu)}
  \,. 
\end{eqnarray}
Here $h^\text{tb}$ is the usual tight-binding hopping matrix.\cite{Santos03}
To save computer time, we employ the checkerboard breakup\cite{LoGu92}
\begin{equation}\label{eq:app:checker}
  e^{\dtau t\sum_{\las i,j\ras} c^\dag_i c^{\phantom{\dag}}_j}
  \approx
  \prod_{\las i,j\ras}
  e^{\dtau t c^\dag_i c^{\phantom{\dag}}_j}
  \,. 
\end{equation}
Using Eq.~(\ref{eq:app:checker}), the numerical effort scales as $N^{2\rD}$ instead
of $N^{3\rD}$, while the error due to this additional approximation is of the same
order $(\dtau)^2$ as the Trotter error in Eq.~(\ref{eq:app:suzuki-trotter}). 

Defining the bosonic and fermionic weights $\wb=e^{-\dtau S_\text{b}}$ and
$\wf=\det(1+\Omega)$ respectively, the partition function finally becomes
\begin{equation}
  \Z_L
  =
  C\,\int\,\D p\,\wb\,\wf\,. 
\end{equation}
One of the advantages of Blankenbecler \etal's\cite{BlScSu81} formalism as well as
the current approach is the close relation to the one-electron Green
function
\begin{equation}
  G_{ij}
  =
  \las \tilde{c}^{\nag}_i \tilde{c}^\dag_j\ras
  +
  \las \tilde{c}^{\dag}_i \tilde{c}^{\nag}_j\ras
  \equiv
  G^{a}_{ij}
  +
  G^{b}_{ij}
  \,. 
\end{equation}
Working in real space and imaginary time, we have\cite{BlScSu81,Hi85}
\begin{equation}\label{eq:app:GA}
  G^{a}_{ij}
  =
  \las \tilde{c}^{\nag}_i \tilde{c}^\dag_j\ras
  =
  (1+\Omega)^{-1}_{ij}\,,
\end{equation}
and
\begin{equation}\label{eq:app:GB}
  G^{b}_{ij}
  =
  \delta_{ij} - G^{a}_{ij}
  =
  (\Omega\,G^{a})_{ji}\,. 
\end{equation}

We would like to mention that despite the formal similarity to the method of
Blankenbecler \etal,\cite{BlScSu81} the numerical realization of the present
approach is quite different. While the grand-canonical method of
Ref.~\onlinecite{BlScSu81} benefits enormously with respect to performance
from a local updating scheme for the phonons, here we use a global updating
in terms of the principal components together with the reweighting method
described in I.  Although this requires us to recalculate the full matrix
$\Omega$ in each MC step, the resulting statistically independent
configurations clearly outweigh the loss in performance, especially for small
$\om_0$ for which autocorrelation times can exceed $10^5$ sweeps when using
the original grand-canonical QMC method.\cite{BlScSu81} The approach proposed
here allows one to perform uncorrelated measurements after each update. 
Moreover, our calculations show that numerical stabilization by means of a
time-consuming singular value decomposition is not necessary for all
parameters considered in Sec.~\ref{sec:results}.  Finally, owing to the phase
factors in the transformed hopping term, all the matrices become
complex-valued. An important practical test in simulations therefore is the
reality of expectation values of observables, \ie, the disappearance of the
imaginary part within statistical errors. 

The phase factors in the hopping term [Eq.~(\ref{eq:app:Hspinless})] give
rise to a minus-sign problem, \ie, $\wf$ can take on negative values, whereas
$\wb$ is always positive. While for a fixed number of electrons the sign
problem diminishes with increasing system size,\cite{HoEvvdL03} the opposite
is true if we fix the electron density. In that case, we find an
exponential decrease of the average sign with increasing cluster size, and a
strong dependence on the band filling (Sec.~\ref{sec:app:sign_results}). 

\subsection{Observables}\label{sec:app:QMC_observables}

The calculation of observables within the formalism presented here is similar
to the standard determinant QMC method.\cite{BlScSu81,Hi85,LoGu92} An important difference is that we
have to use the Lang-Firsov-transformed operators, \ie, 
\begin{equation}
  \las O \ras
  =
  \frac{1}{\Z} \tr(\hat{\tilde{O}}\,e^{-\beta\tilde{\H}})\,. 
\end{equation}
As mentioned above, the MC sampling is based purely on the bosonic
weight $\wb$. This corresponds to a reweighting
of the probability distribution so that the fermionic weight $\wf$ is
treated as part of the observables.\cite{HoEvvdL03} As for a single
electron,\cite{HoEvvdL03} the overlap of the probability distributions
involved is large enough to permit this procedure. Hence, we have
\begin{equation}
  \las O\ras
  =
  \frac{\las O\wf\ras_\text{b}}{\las\wf\ras_\text{b}}
  \,,
\end{equation}
where the expectation value $\las O\ras_\text{b}$ for an equal-time
observable $\hat{O}$ is defined as
\begin{equation}
  \las O \ras_\text{b}
  =
  \frac{\int\,\D p\,\wb \wf \tr_\text{f} (\hat{O} \oB_1\cdots\oB_L)}
  {\int\,\D p\,\wb}
  \,. 
\end{equation}
\subsubsection{Static quantities}

We begin with the electron density
\begin{equation}\label{eq:app:n}
  n
  =
  \frac{1}{N^{\rD}}\sum_i\las \on_i \ras
  \,. 
\end{equation}
The expectation value in Eq.~(\ref{eq:app:n}) can be calculated from the
diagonal elements of the Green function $G^{b}$ [Eq.~(\ref{eq:app:GA})], \ie,
$\las\on_i\ras = \las G^{b}_{ii}\ras$. Similarly, the absolute value of the
electronic kinetic energy per site is given by
\begin{equation}\label{eq:app:Ek}
  \Ek
  =
  \frac{t}{N^{\rD}} \sum_{\las i,j\ras}\las G^{b}_{ji}\ras\,. 
\end{equation}
Equal-time two-particle correlation functions such as
$\rho(\delta)=\sum_i\las\on_i \on_{i+\delta}\ras$ may be calculated with the
help of Wick's Theorem.\cite{BlScSu81,Hi85}

\subsubsection{Dynamic quantities}

An observable of great interest is the time-dependent one-particle Green
function
\begin{equation}\label{eq:app:Gb_tau}
  G^{b}(\bm{k},\tau)
  =
  \las c^{\dag}_{\bm{k}} (\tau)c^{\nag}_{\bm{k}} \ras
  =
  \las e^{\tau\H} c^{\dag}_{\bm{k}} e^{-\tau\H} c^{\nag}_{\bm{k}} \ras
\end{equation}
which is related to the spectral function
\begin{equation}\label{eq:app:akwqmc}
  A(\bm{k},\om-\mu)
  =
  -\frac{1}{\pi}
  \Im G^b(\bm{k},\om-\mu)
\end{equation}
through
\begin{equation}\label{eq:app:maxent}
  G^{b}(\bm{k},\tau)
  =
  \int_{-\infty}^\infty\,d \om\,\frac{e^{-\tau(\om-\mu)}A(\bm{k},\om-\mu)}
  {1 + e^{-\beta(\om-\mu)}}\,. 
\end{equation}
The inversion of the above relation is ill-conditioned and requires the use
of the MEM.\cite{JaGu96} Fourier transformation leads to
\begin{equation}
  G^{b}(\bm{k},\tau)
  =
  \frac{1}{N}\sum_{ij}  e^{\rmi \bm{k}\cdot(\bm{r}_i-\bm{r}_j)} G^b_{ij}(\tau)\,. 
\end{equation}
The allowed imaginary times are $\tau_l=l\dtau$, with nonnegative
integers $0\leq l\leq L$. Within the QMC approach, we have\cite{BlScSu81,Hi85}
\begin{equation}\label{eq:app:Gb_tau_QMC}
  G^{b}_{ij}(\tau_l)
  =
  (
   G^{a} B_1\cdots B_l
  )_{ji}
    \,. 
\end{equation}
Another interesting quantity is the one-electron density of states
\begin{equation}\label{eq:app:dosgen}
  \rho(\om-\mu)
  =
  -\frac{1}{\pi} \Im G(\om-\mu)
  \,,
\end{equation}
where $G(\om-\mu)=N^{-1}\sum_{\bm{k}} G(\bm{k},\om-\mu)$. It may be obtained numerically
via
\begin{equation}\label{eq:app:DOS}
  \rho(\tau)
  =
  \frac{1}{N}\sum_i G^{b}_{ii}(\tau)\,,
\end{equation}
and subsequent use of the MEM. 

\begin{figure}[t]
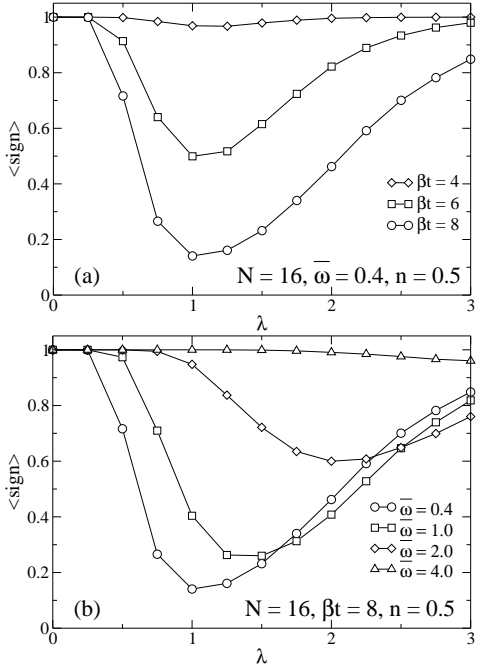

  \includegraphics[width=0.35\textwidth]{F12a.eps}\\
  \includegraphics[width=0.35\textwidth]{F12b.eps}
\caption{\label{fig:app:sign_beta_omega}
  Average sign $\sign$ [Eq.~(\ref{eq:app:sign})] of the fermionic weight
  $\wf$ as a function of electron-phonon coupling $\lambda$ in one dimension
  (a) for different inverse temperatures $\beta$, and (b) for different
  values of the adiabatic ratio $\omb$.  Here and in subsequent figures
  $\dtau=0.05$, lines are guides to the eye only, and errorbars are suppressed if
  smaller than the symbols shown.} 
\end{figure}
%

\subsection{Sign problem}\label{sec:app:sign_results}

Since a possible sign problem crucially affects QMC simulations, this section
is devoted to a detailed investigation of the dependence of the average sign
on the various parameters. 

While Hamiltonian~(\ref{eq:app:Hspinless}) is symmetric with respect to a
particle-hole transformation for $\mu=-\Ep$ (half filling), this symmetry is
broken if we use the checkerboard approximation~(\ref{eq:app:checker}), so
that $n\neq0.5$ for the above choice of the chemical potential. To simplify
calculations, the results for the average sign at $n=0.5$ presented below
have therefore been obtained using the exact hopping term. For general band
fillings, we find that the checkerboard breakup requires different values of
$\mu$ to perform simulations at the same electron densities, but results are
identical within statistical errors if $\mu$ is adjusted accordingly. 

The derivation of the QMC algorithm in Sec.~\ref{sec:app:QMC_partition} is
valid for any dimension D of the lattice under consideration. Nevertheless,
here we only report results for the sign problem in $\rD=1$, the case
considered in Sec.~\ref{sec:results}, and make some remarks about the
influence of the dimensionality at the end. 

The average sign is defined as
\begin{equation}\label{eq:app:sign}
  \sign
  =
  \frac{\las\wf\ras_\text{b}}{\las|\wf|\ras_\text{b}}
  \,,
\end{equation}
with the fermionic weight in the present case given by $\wf=\det(1+\Omega)$
[Eq.~(\ref{eq:app:omega})]. We begin with the dependence of $\sign$ on the
electron-phonon coupling strength. For simplicity, we show results for
$n=0.5$, while the effect of band filling will be discussed later. The choice
$n=0.5$ is convenient since we know the chemical potential, and we shall see
below that the sign problem is most pronounced for a half-filled band. 
Moreover, all existing QMC results for the spinless Holstein model are for
half filling, so that it is interesting to see how the sign problem affects
simulations within the current approach. 

\begin{figure}[t]
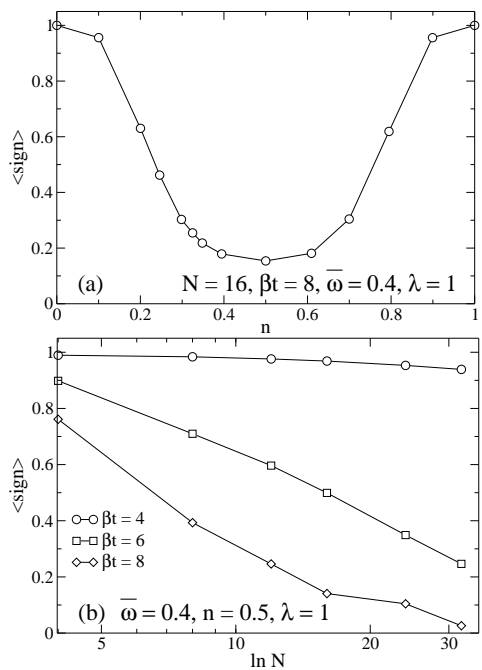

  \centering
  \includegraphics[width=0.35\textwidth]{F13a.eps}\\
  \includegraphics[width=0.35\textwidth]{F13b.eps}
\caption{\label{fig:app:sign_n_N}
  Average sign as a function of (a) band filling $n$, and (b) system
  size $N$.} 
\end{figure}

Figure~\ref{fig:app:sign_beta_omega}(a) reveals that the average sign takes
on a minimum near the point where---depending on the filling---the cross over
to small polarons or a CDW state occurs. In the adiabatic regime $\omb<1$,
the cross over condition is $\lambda>1$
[Fig.~\ref{fig:app:sign_beta_omega}(a)], whereas for $\omb>1$ we have $g>1$
[Fig.~\ref{fig:app:sign_beta_omega}(b)]. At weak and strong coupling, the
average sign is close to 1, so that accurate simulations can be carried out. 
As shown in Fig.~\ref{fig:app:sign_beta_omega}(b), we find that $\sign$
becomes very small for $\omb<1$ and intermediate $\lambda$, while it
increases noticeably in the nonadiabatic regime $\omb>1$. Owing to the
absence of any autocorrelations in our approach, the present method therefore
represents a significant improvement of existing algorithms, as the latter
face severe autocorrelations even for $\omb\gtrsim1$, thereby limiting the
temperature range and cluster size. 

As illustrated in Fig.~\ref{fig:app:sign_n_N}(a), the average sign depends
strongly on the band filling $n$. While $\sign\approx1$ in the vicinity of
$n=0$ or $n=1$, a significant reduction is visible near half filling $n=0.5$. 
The minimum occurs at $n=0.5$, and the results display particle-hole symmetry
as expected. Here we have chosen $\beta t=8$, $\omb=0.4$ and $\lambda=1$, for
which the sign problem is most noticeable according to
Fig.~\ref{fig:app:sign_beta_omega}. Note that there is no sign problem in
simulations for, \eg, the Hubbard model at half filling, where the
(fermionic) weights of the $\UP$ and $\DO$ electrons are equal and their
product---entering the partition function---is positive.\cite{wvl1992} This can also be
expected within the current approach if we consider the half--filled spinfull
Holstein or Holstein-Hubbard model, and work along this line is in progress. 

Finally, in Fig.~\ref{fig:app:sign_n_N}(b), we report the average sign as a
function of system size, again for $n=0.5$. The behavior is strikingly
different from the one-electron case considered in I.  While in the latter
$\sign\rightarrow1$ as $N\rightarrow\infty$, here the average sign decreases
nearly exponentially with increasing system size. Obviously, this limits the
applicability of our method. However, as illustrated in
Sec.~\ref{sec:results}, we can nevertheless obtain accurate results at low
temperatures, small phonon frequencies, and over a large range of the
electron-phonon interaction. Moreover, we would like to point out that for
such parameters, existing methods suffer strongly from autocorrelations,
rendering simulations extremely difficult. 

The dependence of the sign problem on the dimension of the system is similar
to the single-electron case.\cite{HoEvvdL03} For the same parameters $N$,
$\omb$, $\beta t$ and $\lambda$, the minimum in $\sign$ at intermediate
$\lambda$ becomes more pronounced as one increases the dimension of the
cluster. 

To conclude with, we would like to point out that, in principle, the sign
problem can be compensated by performing sufficiently long QMC runs, although
we have to keep in mind that the statistical errors increase proportional to
$\sign^{-2}$ (Ref.~\onlinecite{wvl1992}). Owing to the purely phononic
updates, the present algorithm is very fast and we have made up to about
$2.6\times10^7$ single measurements to obtain the results presented in
Sec.~\ref{sec:results}. The use of the reweighting method requires us to
store the measured values of observables in every MC step, and
to perform the Jackknife error analysis\cite{DavHin} at the end of the run. 
For example, in the case of the Green function~(\ref{eq:app:Gb_tau}), $L+1$
complex numbers have to be stored for each momentum $k$ in every step. 
Depending on the observables of interest, this sets a practical limit for the
maximal number of measurements due to restrictions in available disk space.




\end{document}